\documentclass[fleqn,useAMS,usenatbib]{mnras}

\usepackage{newtxtext}

\usepackage[T1]{fontenc}
\usepackage{ae,aecompl}
\usepackage{color}
\usepackage[english]{babel}
\usepackage{graphicx}
\usepackage{amsmath}
\usepackage{amssymb}
\usepackage{bm}
\usepackage{ulem}

\begin{document}

\newcommand{\dd}{\mathrm{d}}
\newcommand{\rhob}{\rho_\text{b}}
\newcommand{\Tb}{T_\text{b}}
\newcommand{\kB}{\mbox{$k_\text{B}$}}
\newcommand{\rate}{\mbox{erg cm$^{-3}$ s$^{-1}$}}
\newcommand{\gcc}{\mbox{g~cm$^{-3}$}}
\newcommand{\xr}{x_\text{r}}
\newcommand{\gs}{g_\text{s}}
\newcommand{\me}{m_\text{e}}
\newcommand{\msun}{\mbox{$M_\odot$}}

\def\la{\;\raise0.3ex\hbox{$<$\kern-0.75em\raise-1.1ex\hbox{$\sim$}}\;}
\def\ga{\;\raise0.3ex\hbox{$>$\kern-0.75em\raise-1.1ex\hbox{$\sim$}}\;}


%

\title[Search for a quasi-periodic structure]{Search for a possible quasi-periodic structure based on data
of the SDSS  DR12 LOWZ}

\author[A.~I.~Ryabinkov, A.~D.~Kaminker]
{A.~I.~Ryabinkov, 
A.~D.~Kaminker \\ 
Ioffe Institute,
Politekhnicheskaya 26, 194021 St.~Petersburg, Russia  \\ 
e-mail: ryabin60@gmail.com, kam.astro@mail.ioffe.ru
}

\date{Accepted 2022 xxxx. Received 2022 xxxx; 
in original form 2022 xxxx}


\maketitle    
       
\begin{abstract}
We  carry  out  a
statistical analysis  of   the
spatial distribution of galaxies
at cosmological redshifts
$0.16 \leq z \leq 0.47$
based  on
the  SDSS\   DR12\  LOWZ   catalogue.
Our aim is to search and study possible
large-scale quasi-regular structures
embedded in the  {\it cosmic web}.
We calculate
projections  of the Cartesian
galaxy coordinates
on different axes (directions)
densely covering certain regions in the sky
to  look for  special directions
along which one-dimensional distributions
of the projections
contain significant  quasi-periodic  components.
These components appear
as peaks in the  power spectra and
lie in a narrow range of
wave numbers $0.05 < k < 0.07$.
Particular attention is paid to the evaluation
of the significance of the  peaks.
It  is  found
that  the  significance  of  the dominant peaks
for   some selected directions  exceeds  $(4 - 5)\sigma$.
In order to reduce  possible  selection effects,
we create a  mock  homogeneous  catalogue  of
spatial distribution of galaxies by
adding a random set of artificial objects (points)
to  the  real  galaxies  under study.
The power spectrum of this cumulative
model  data  also demonstrates
significant peak corresponding to approximately
the same scale. As a result
we assume the existence of an anisotropic
cosmological quasi-periodic
structure  with characteristic scale
$(116 \pm 10)~h^{-1}$~Mpc.
\end{abstract}


%
\section{INTRODUCTION}
\label{sec:intro}
There are observational evidences 
in literature (e.g. \citealt{saar02}, \citealt{einast14}
and references therein)  
that some domains of the 
spatial distribution of   
cosmologically distant objects  show 
elements of spatial  regularity   
with  a rather wide range of
large scales   $~ (110 - 140)~h^{-1}$~Mpc. 
Such traces of regularity can be   
scattered at random locations
in a network  formed by galaxies, galaxy clusters and superclusters
representing both high-density regions  --
walls,   filaments  and nodes,   and   low-density  
regions  -- giant  voids,  occupying the bulk of the space in the Universe  
(for review  see, e.g.  \citealt{weygaert09}, 
 \citealt{einast14}, \citealt{weygaert16}).

Mention can  be made here   of the    
deep   $[ z \sim (0.1 - 0.4) ]$   pencil-beam 
surveys of galaxies
in the direction of the North and South galactic poles
produced  by \citet{beks90}  and continued by  
\citet{szetal91}, \citet{szetal93} and \citet{koo93},
which  showed that the 1D distribution   of 
galaxies may exhibit quasi-periodic components
with characteristic scale $\sim$ 130~$h^{-1}$~Mpc.
The heated discussion in the literature about the significance and meaning 
of these results  lasted  for more than a  decade.
\citet{kp91}  showed that the quasi-regularity found in the 
pencil-beam  {galaxy samples} can be explained 
within   the  assumption of 
a random arrangement of galaxy clumps. 
Only much later, simulations carried out by \citet{yoshida01}, 
taking into account the most probable cosmological models, 
showed that the probability of a random origin of the 
detected quasi-regularity can be quite small (less than $10^{-3}$). 
That indirectly confirmed the 
quite noticeable
significance 
of the discussed quasi-regularity,  
but the issue is still open.

The conjecture about   the existence of traces of regularity 
in the distribution of galaxies was supported 
by an analysis  of  \citet{landy96},
based on the Las Campanas Redshift Survey  data (at $z \la 0.1$). 
They plotted 2D Fourier transforms  
of the galaxy distributions
in six thin slices, three in each 
the North and South Galactic  hemispheres,  and found significant peaks 
in certain directions of the  wave vector ${\bf k}$
in the 2D power spectra. The peaks correspond  
to  quasi-periods   of  $\sim 100~h^{-1}$~Mpc.

An important advancement of the hypothesis of
regularity in the distribution  of
cosmologically distant objects were
the works of  J. Einasto 
and co-authors (e.g. \citealt{einastM94,  einast97a, einast97b, 
einast97c, nat_einast97},  and  \citealt{einastM16_shell}, see also \citealt{kerscher98}). 
It was shown that the cosmological   
network formed by rich  galaxy clusters and
superclusters  and voids between them may show 
traces of a regular spatial (cubic-like or shell-like) structure, 
with characteristic scales  $(115 - 140)~h^{-1}$~Mpc.

Somewhat  later  \citet{saar02} proposed   a
new method for determining    quasi-regularity  of
a cubic-like lattice formed by a network  
of superclusters and voids. 
It was shown that a significant quasi-periodicity with 
a characteristic scales 
$~(120 - 140)~h^{-1}$~Mpc can be observed  along 
certain directions in space. 

Additionally,  \citet{einast11},  \citet{einast11a},  and   \citet{einast19}
(also references therein) 
considered  the effects of  spatial phases  synchronization  
related to perturbations  in  different  spatial  scales  
and their impact on the formation 
of the observed cosmic web.  
It seems to  us  that  such 
effects of phase ordering   
can give additional   motivation  to  the search
for traces of regularity on larger scales.

In the present work, 
we try to find  further confirmations of this hypothesis, 
but here we consider much larger redshifts ($0.16 \leq z \leq 0.47$) 
than  it was performed in the works cited above.
We  focus   on studying  
the spatial distribution  of galaxies 
in the northern hemisphere 
based on the data of SDSS DR12 LOWZ catalogue
(e.g.  \citealt{dawson13},\  \citealt{reid16}).
In  our statistical analysis  we use
methods, which in a sense can be attributed to methods of  integral  
geometry, and combine them with calculations of the power spectra 
of the obtained distributions.

It   is   appropriate to mention  that methods of integral geometry, 
in particular the method of Minkowski functionals 
(e.g.  \citealt{kerscher97},  \citealt{sahni98},  \citealt{kerscher00},  
\citealt{mecke00},   \citealt{appbuch22}), 
have been widely used to study the morphology and structure
of cosmologically distant  matter   in the Universe
during a few past decades.
In our case we use  the   so  called  
{\it Radon transform} (e.g.\   \citealt{deans07},  \citealt{starcketal05})
and  some of  its modification (see Sect.~\ref{sec:cub})  
to look for 
traces of regularity and anisotropy 
in   the  spatial distribution of galaxies. 
The methods  used are  not  fully integral 
since  they  retain   selected directions in comoving 3D space
on which all the Cartesian coordinates of   the galaxies  
belonging to a  sample are projected (integrated).  
As a result, 
a one-dimensional distribution of  projections of the 
Cartesian coordinates   of galaxies  
is constructed for any fixed direction $X$ 
and  then  the 1D power spectrum  
of  this  distribution
is calculated.
This  procedure allows  to determine those directions
in space in which a
significant quasi-periodicity can be detected.
The  method   turns  out  to  be  quite  sensitive to very weak  
traces  of  ordering  in the  large-scale structure.

In our previous  papers  
(e.g.  \citealt{rkk13}, \citealt{rk14}, \citealt{rk19}, hereafter Paper~I,
\citealt{rk21}, hereafter Paper~II)
we  analysed  mainly   the   so called 
{\it radial distributions} of  cosmological objects
(the  luminous  red  galaxies, LRGs, or  the brightest cluster galaxies, BCGs),   
which can be  imagined  as  homogeneous  spherical  shells  
filled  with  the  objects 
depending only on the comoving distance from the center of
spherical system, while the angular coordinates of  the  objects
are completely ignored.  
In  these  papers, 
the power spectra 
of the radial distributions $P_R(k)$  were  calculated  and significance levels
of peaks in the  region  $k \sim (0.05 - 0.07)~h~{\rm Mpc}^{-1}$
were estimated in some way or another.   
So in  Paper~I   the  method   to  assess   the significance 
of the peaks was proposed
based on the exponential distribution 
of peak amplitudes (see   Eq.~(\ref{calF}))
with  taking into account
dependences  of  $\langle P_R(k) \rangle$ on $k$, 
where $\langle ... \rangle$ is  averaging over 
the  ensemble of realizations. 
It was shown, in particular,
that  $\langle P_R(k) \rangle$ for extremely large ensembles
of radial distributions constructed  with respect to 
various centers  scattered  over  the homogeneous  space
tends to be equal to  3D power spectrum $P_{3D}(k)$
averaged over various directions of wave vector ${\bf k}$.  
Using the method  proposed in Paper~I, 
the results of  \citet{rkk13} and  \citet{rk14} were reevaluated 
and it turned out that the significance of the peaks 
in the radial power spectra obtained in these articles 
did not exceed   $3 \sigma$  level
contrary to the estimates made in these works.

In  Paper~II   on  the  basis  of  several  samples 
extracted from the SDSS DR7 
catalogue by \citet{kazetal10} 
we  considered  
the radial distributions of  the  LRGs,
within  an  interval $0.16 \leq z \leq 0.47$, 
restricted  by  rectangular  regions  ({\it sectors}) in the sky. 
It was shown that the radial distributions 
in some selected sectors 
characterized by certain directions 
contain significant    $(\ga  4 - 5 ~ \sigma)$    
quasi-periodic  components with the same periods 
($\ga 110~h^{-1}~{\rm Mpc}$)  as
the values  received in  \citet{rkk13} and  \citet{rk14}
for a much wider view of the sky.
Perhaps this  indicated the anisotropy of the quasi-regular
structure  distributed over a certain spatial   region.

Also in Paper~II we  proposed  to apply the  Radon transform
and subsequent calculations  of 1D power spectra
for searching   the direction of the maximum peak amplitude. 
For various samples under consideration
it   was 
found  that   there is  quite   a  narrow bunch of directions
exhibiting   significant quasi-periodicities 
with close periods   $116 \pm 10~h^{-1}~{\rm Mpc}$.
Such directions are 
characterized by the angular coordinates 
of the  Equatorial coordinate system 
clustering  around  
right ascensions 
$\alpha_0 \simeq 175^\circ \pm 2^\circ$
and declinations 
$\delta_0 \simeq  25^\circ \pm 2^\circ$. 
In addition, in  Paper~II  we  carried out  preliminary 
calculations based on the extended 
SDSS  DR12  data  in a wider range ($0.16 \leq z \leq 0.72$) 
and  obtained   confirmation of the possible existence 
of an anisotropic quasi-regular structure
with the same characteristic scale.

In the present study we confirm our previous conclusions 
about the possible existence 
of a   weak \   (i.e. \  manifested  by the integral methods  
mentioned  above)  
anisotropic quasi-periodic  structure 
in the spatial distribution of galaxies.
Moreover, we find approximately the same specific  
direction of the $X$-axis,   
the projection onto which 
the Cartesian coordinates
of galaxies 
contains a quasi-periodic component 
corresponding to the highest level of 
significance ($\ga 4 - 5 \sigma$).
These findings encourage us to continue similar
statistical analysis on larger samples  beyond   the   SDSS data.
The first very preliminary experience  with catalogs 
of photometric data   is represented   in  Sect.~\ref{sec:cd}.

In Sect.~\ref{sec:data} we describe the observational data,
employed  in this work.
In Sect.~\ref{sec:bd} we determine  
basic quantities and definitions used 
in further  analysis.
In Sect.~\ref{sec:rectang} we  
employ  the
Cartesian coordinate system (CS)
and introduce  the Radon transform with
its  subsequent 1D  power spectrum calculations. 
In Sect.~\ref{sec:cub} 
we introduce and apply another technique, 
the  so-called  oriented cuboids method. 
In Sect.~\ref{sec:mod} we show that the sample of   the   LOWZ  galaxies
within the interval   $0.16 \leq z \leq 0.4$  
 in the strict sense is not homogeneous
(not volume-limited) and build 
(somewhat artificially)
a model homogeneous 
sample. 
Conclusions and discussions 
of the results are  given in Sect.~\ref{sec:cd}.

\section{Data used}
\label{sec:data}
In the present  consideration  we use data on galaxies 
with redshifts $0.16  \leq z \leq 0.47$ 
from the  SDSS  DR12  (LOWZ  and CMASS)  catalogues   
accumulated only for the northern hemisphere in the sky and represented
in   three  data  files: \\
$galaxy$\_$DR12v5$\_$LOWZ$\_$North.fits.gz$, \\
$galaxy$\_$DR12v5$\_$CMASSLOWZE3$\_$North.fits.gz$ \\
$galaxy$\_$DR12v5$\_$CMASS$\_$North.fits.gz$, \\
which are available in the Science Archive Server.%
\footnote{https://data.sdss3.org/sas/dr12/boss/lss/}
A description of  the  catalogues   of  the  DR12  can be found,
e.g. in   \citet{dawson13},   \citet{alam15},  \citet{reid16}.

We restrict ourselves mainly by  spectroscopic   data 
on redshifts  (with  exceptions in  
Sect.~\ref{sec:mod} and partly in  Sect.~\ref{sec:cd})
with  redshift uncertainty $\Delta z /(1+z)  \sim  5 \times 10^{-4}$
(see, e.g. \citealt{bolton12}),
which provide a reliable determination 
of  heterogeneity,  both irregular and regular,  
in the considered scales. 
However, our first experience in the statistical analysis
of  galaxy cluster  catalogues  
(e.g.   \citealt{whl12} and \citealt{wh22}) 
with photometric redshifts,  determined with
uncertainty  $\sim  (0.01 - 0.02)$ at $z \la 0.5$, 
confirmed  to some extent  the findings
obtained from the spectral data.
This will be discussed briefly in the Sect.~\ref{sec:cd}. 

\begin{figure}     
\includegraphics[width=0.45\textwidth]{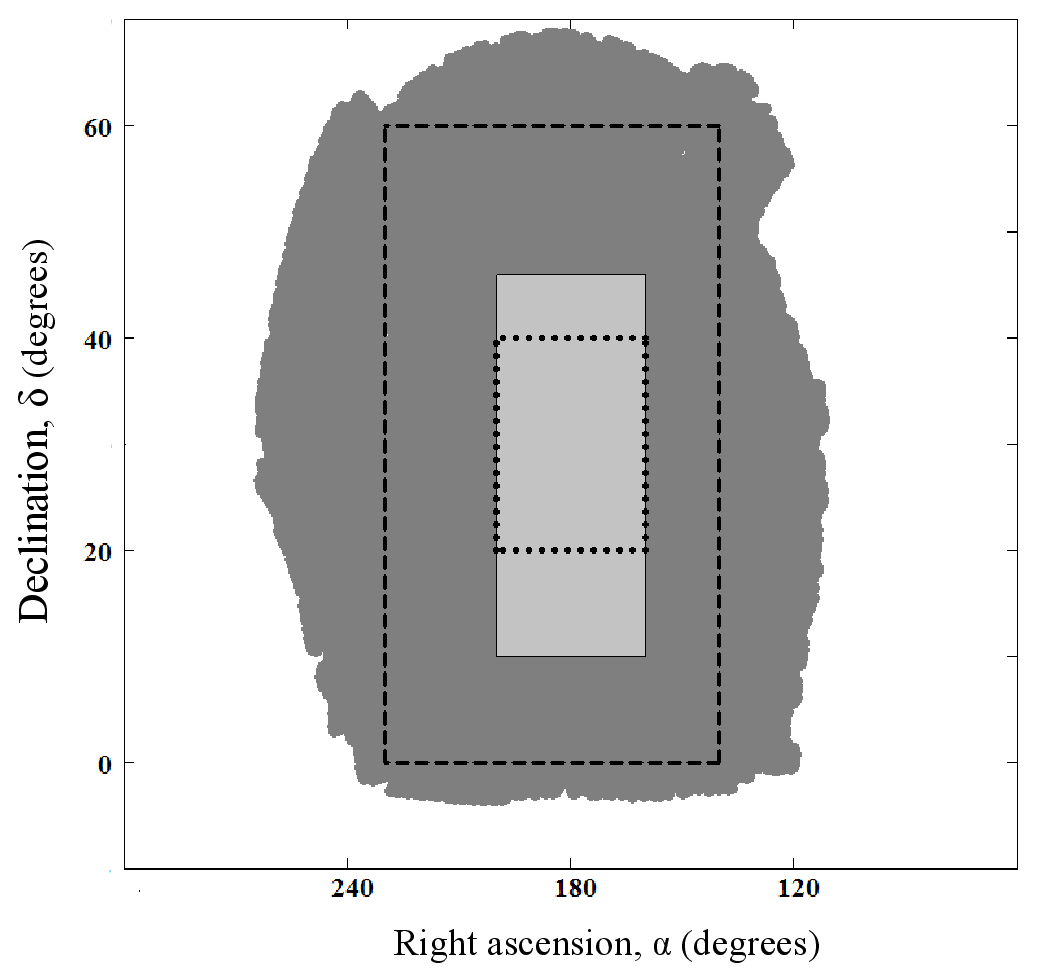}%
\caption{
(Colour online) 
Angular distribution of  galaxies 
over the sky from the  SDSS\   DR12\  LOWZ\   data 
in the Equatorial  CS; 
the entire  grey  coloured  region
together with the rectangle  included in it
(marked in light grey)   
comprise  the  LOWZ   northern  hemisphere  sample
with a somewhat artificially filled regions (gaps) 
of data incompleteness (see text);
{\it dashed} lines delineate the general rectangular region of a given
statistical consideration 
with $140^\circ \leq \alpha \leq 230^\circ$
and $0^\circ \leq \delta \leq 60^\circ$,
$\alpha$   and $\delta$  are  indicated  near  the  coordinate axes   
( $\alpha$ is shown
in a  standard way:   west  on  the  right,  east  on  the left). 
Light grey marks a smaller rectangular region on the celestial sphere, 
$160^\circ \leq \alpha \leq 200^\circ$  and 
$10^\circ \leq \delta \leq 46^\circ$,   
in which the effects considered  in the text
are especially significant.  The dotted lines  bound  a 
smaller  area of the rotation of   $X$-axis
described in  Sect.~\protect{\ref{sec:rectang}}.
}  
\label{f:rect}
\end{figure}

The SDSS  DR12  LOWZ  region  
is    shown in Fig.~\ref{f:rect}
in  the  Equatorial  coordinates
mainly in the northern hemisphere 
(marked in grey).
Our basic  sample includes galaxies,
falling into  a large rectangle
region delineated 
by  thick dashed lines. This  helps  to avoid
possible effects of the irregular edges 
on the borders of the entire grey area.
We choose  the same intervals 
of the right ascension 
$140^\circ \leq \alpha \leq 230^\circ$
and declination $0^\circ \leq \delta \leq 60^\circ$ 
as  it was  chosen 
in  Paper~II  for the SDSS DR7 catalogue.

It should be noted that  the
selected rectangular area
includes  parts of  two  gaps
(filled in fig.~\ref{f:rect}, see below)
associated   with  the  incompleteness of the sample of galaxies
in the LOWZ  catalogue. The gaps  correspond to
earlier observations
with a different galaxy selection algorithm
(see  \citealt{reid16} for details).
Inside a rectangular area
the main gap with a missing number of galaxies
located within the  area in the sky
$185^\circ \la \alpha \leq  230^\circ$
and  $25^\circ \la \delta \la 40^\circ$;
the second gap at $\delta \ga 0$ is not statistically significant
(Figures~8 and A1 of  \citealt{reid16}).

\citet{reid16} generated a sample `LOWZE3'
including all areas of incompleteness  that overlap
with the rectangle represented in Fig. 1 ({\it chunks 3-6}
in the terminology of \citealt{reid16}).  
Mentioned above the CMASS catalogue  was combined with 
the LOWZE3 sample to create the `CMASSLOWZE3' sample 
which is also available on the DR12  Science Archive Server.
To fill the gaps inside the rectangular region and 
in adjacent areas shaded in solid  grey in Fig. 1,
we had to extract  the `LOWZE3'   
from   the  `CMASSLOWZE3' (subtracting the CMASS  galaxies) 
and clear the resulting sample from galaxies  
which  are also present in the LOWZ catalogue.
Then we  created a new combined 
sample  `LOWZ'  + (refined) `LOWZE3'
which  is  shown  in  Fig.~\ref{f:rect}.%
\footnote{
The sample `LOWZ'+`LOWZE3' have been
also produced by \citet{reid16} but the corresponding file
is not available in the Science Archive Server
indicated above.}
The sample selected in such a way contains 176,962 galaxies
detected within  the  large  rectangle   in Fig.~\ref{f:rect} 
and  the redshift  interval   indicated above.
In  Sects.~\ref{sec:rectang} and \ref{sec:cub}  we   briefly
discuss  that the results obtained turned out to be 
robust to such  (not quite  strict) a gap-filling   procedure.

Following  methodology described in   Paper~II
we scan the selected  rectangle region  
using a mobile trial sector with angular dimensions
$5^\circ$ (right ascension) and $25^\circ$ (declination).
In this way, we select  144  (partly overlapping) 
sectors  for  the whole rectangular area. Within each sector we
build   the   so-called radial (1D) distribution function  of galaxies
$N_R(D)$, where $D$ is defined in Eq.~(\ref{D}),%
\footnote{
The radial (shell-like) distribution function $N_R(D)$,
averaged  over angles  $\alpha$
and  $\delta$  within 
a  selected  area (e.g.  sector) in the sky, 
was described  and analysed in Papers~I and  ~II.} 
and calculate the 1D radial power spectra $P_R (k)$
for   the   distributions.

Thus,  six sectors were found in Paper~II 
on  the base of  SDSS DR7 data, 
in which the significance of peaks in the radial power spectra 
within the interval 
$0.05 < k < 0.07~h~{\rm Mpc}^{-1}$
exceeded  $3 \sigma$. 
In  the  present  analysis   
we act   differently, 
we  choose  a  smaller rectangular  region
with $160^\circ \leq \alpha \leq 200^\circ$
and $10^\circ \leq \delta \leq 46^\circ$
(marked in light grey in  Fig.~\ref{f:rect}), 
including  three (central) 
of the six  indicated  sectors  with the 
highest   significance levels of  the 
peaks  in the corresponding 
radial power spectra.

In Sect.~\ref{sec:rectang}\  (Fig.~\ref{f:PXk})
and in Sect.~\ref{sec:mod}\  (Fig.~\ref{f:model}) 
we consider
only this (smaller) rectangular area
in which   we can  expect  a higher probability 
to detect traces of quasi-periodicity
using  the approach  described  in these sections.

Note   that the data of   DR12\   LOWZ  
catalog  that we use here
are  not  strictly  homogeneous 
[see, e.g. Fig.~\ref{f:model} (a)], so
in  Sect.~\ref{sec:mod} 
we use an extended model sample, 
specially  prepared  as homogeneous one.
The results turned out to be resistant to such 
simulations.

\section{Cartesian Coordinate System.  Basic definitions}
\label{sec:bd}  

Let us consider   spatial  distribution of galaxies  presented 
in the   DR12  LOWZ  catalogue  within
certain   region in the sky   using  the  Cartesian CS:
\begin{eqnarray}
\label{XYZ}
&  & X_i = D (z_i) \sin(90^\circ - \delta_i) \cos \alpha_i       \\
\nonumber  
&  & Y_i = D (z_i)  \sin(90^\circ - \delta_i) \sin \alpha_i          \\
\nonumber  
&  &  Z_i = D (z_i) \cos(90^\circ - \delta_i),
\end{eqnarray}
where $D (z_i)$ is radial comoving distance of  
$i$-th galaxies  with redshifts $z_i$, 
measured in  $h^{-1}~{\rm Mpc}$    
(e.g. \citealt{khs97};  \citealt{h99})
\begin{equation}
D (z_i) = {c \over H_0}\, \int_0^{z_i} 
{1 \over \sqrt{\Omega_{\rm m} (1+z)^3 + 
\Omega_\Lambda}}\ {\rm d}z, 
\label{D}
\end{equation}
$H_0=100~h$~km~s$^{-1}$~Mpc$^{-1}$ is the present Hubble constant,
$c$ is the speed of light; 
$\alpha_i$ -- its right ascension and $\delta_i$ -- declination;
in both the coordinate systems  an observer is at 
the  origin.   

Hereafter    (for comparison with our previous results),
we use the  same $\Lambda$CDM model
with  $\Omega_{\rm m}=0.25$ and 
$\Omega_{\Lambda}=1-\Omega_{\rm m}=0.75$
as  it  is chosen  in  Paper~II.   

Following  Paper~II  we apply the method  
of constructing 1D distributions of projections of   the 
Cartesian  coordinates of galaxies  
onto different $X$-axes rotated relative to each other
around  the  origin.  
In this approach the basic value for  the  spectral  analysis  
is  an  1D  distribution function $N_X(X)$
collecting all projections of the coordinates 
within outlined angular region
on the  fixed  X-axes;  
$N_X (X) {\rm d}X$ is a number of galaxies
inside an interval ${\rm d} X$. 
As in Paper~II  we use the binning approach%
\footnote{The $X$-axis is divided into a finite number of non-overlapping 
intervals - bins, each of which accumulates a certain amount 
projections of  coordinates of galaxies.  
Thus the continuous distribution of projections  $N_X(X)$
is replaced by a discrete one $N_X(X_c^l)$ (see Eq.~\ref{NNX}).} 
and calculate so-called 
normalized 1D distribution function in the comoving CS as a number
of  projections of the Cartesian coordinates inside 
longitudinal non-overlapping bins:
\begin{equation}
{\rm NN} (X_c^l) = {N_X (X_c^l) - S_X
\over \sqrt{S_X}},
\label{NNX}
\end{equation} 
where  $X_c^l$   is  a  central  point of  a  bin, 
$l=1,2,...\ {\cal N}_b$ is   numeration of  bins,
$S_X$ is  a mean value of  the 1D  distribution 
$N_X (X_c^l)$ over all  bins  under study. 
Essentially, the magnitude of  NN$(X_c^l)$ 
can be considered as 
a function of the signal-to-noise ratio along the selected 
$X$-axis. We  use  such a representation of  Eq.~(\ref{NNX})
for evaluations in  Sect.~\ref{sec:cd}.

The values of   NN$(X_c^l)$ allows one to calculate 
corresponding 1D power spectrum
\begin{eqnarray}
\label{PXk}
P_X(k_m) & = & |F_X^{1D} (k_m) |^2 = \\
\nonumber 
&  & {1 \over {\cal N}_b} \left\{ \left[ \sum_{\ l=1}^{{\cal N}_b} 
NN(X_c^l) \cos (k_m X_c^l) \right]^2  \right. \\
\nonumber
& + & 
\left. \left[ \sum_{\ l=1}^{{\cal N}_b} NN(X_c^l) \sin(k_m X_c^l) \right]^2 \right\},
\end{eqnarray}
where  $F_X^{1D} (k_m) = 
({\cal N}_b)^{-1/2} \sum_{\ l=1}^{{\cal N}_b} NN(X_c^l)\ e^{-ik_m X_c^l}$
is the one-dimensional discrete Fourier transform,  
$k_m =2 \pi m / L_X$ is  a 
wavenumber corresponding to 
an integer harmonic number $m=1,\ 2,\ ..., {\cal M}$,\
${\cal M}=\lfloor {\cal N}_b/2 \rfloor$  is
a maximal number (the Nyquist number) 
of independent discrete harmonics, $\lfloor x \rfloor$ denotes 
the greatest integer  $\leq x$,  $x$ is an arbitrary real (positive) number; 
$L_X$  is the whole interval 
in the configuration space, i.e.   the   so-called  {\it sampling length}. 

Then we rotate the coordinates $XYZ$  at the certain Euler angles
so that the new $X'$%
\footnote{Hereafter, denotation $X'$ instead of general denotation $X$
indicates the axis $X$ of  the rotating  CS.}
is oriented in a certain direction ($\alpha'$ and $\delta'$) 
relative to  the  initial Equatorial CS. Performing a sequence of such 
rotations we search for  $X_0$-axis  or a direction along
which the 1D power spectrum displays the most significant
peak for $k$ within the interval $0.05 < k  < 0.07$.

To control the uniformity of statistics for different directions of $X$
we fix the same boundaries of the rotated axes 
$464 \leq  X  \leq  1274~h^{-1}$~Mpc
which  contain ${\cal N}_b = 81$ independent bins with  
a width $\Delta_{X} = 10~h^{-1}$~Mpc. 
On the other hand,  we  
limit  ourselves  to such 
a scanning region of  $X'$ directions
that on the boundary of this  region
all  bins  $\Delta_{X'}$  would be  filled
(more or less evenly)  with  projections of
galaxy  coordinates.
In addition, we need to make sure that the amplitudes of the peaks
of power spectra in the neighborhood of $k$ indicated above,
calculated for borders of
this region would be small enough
(significance  should  be  $\la 3\sigma$).  
The latter  condition 
provides us by the background area
around the domain  with increased peak amplitudes  
where the peaks are likely to have random nature.
Such  areas  are  suitable for
assessing  the significance of the peaks (see below).

To satisfy these conditions, we restrict 
the area of analyzed directions of  $X'$ by intervals
$160^\circ \leq \alpha' \leq 200^\circ$ and 
$20^\circ \leq \delta' \leq 40^\circ$, i.e.  the smallest
rectangle in Fig.~\ref{f:rect}  bounded by dotted lines.%
\footnote{Not confuse  with the region that provides 
the analyzed sample of galaxies, i.e. 
the entire light  grey rectangular
region in Fig.~\ref{f:rect}.}

Actually, here we deal with
a  discrete  analog  of   the  so-called  3D Radon transform 
(e.g.  \citealt{deans07})  
applied to  selected  data  in comoving  coordinate system, 
i.e. we  summarize   the projections of all subsample points,
falling into each bin along  $X$.
Thereafter, we  exploit  two main properties of the Radon transform
(i) {\it translation invariance} that allows one  to transfer the projections 
of  the Cartesian coordinates of galaxies on the given  $X$ axis
to another $\hat{X}$ axis  parallel to the original  one,
(ii)  {\it linearity}, which allows one  to  
summarize the  projections obtained for individual sectors 
in the sky in the total sum of  projections
to get  a single Radon transform 
for  the entire  sample.

It was  verified  numerically in  Paper~I   for  a  set  of
the  power spectra $P_R(k)$  simulated  specially
for  radial  distributions  of  objects  ( Sect.~3 in Paper~I) 
that the cumulative  probability  function of random
peak  amplitudes  $P_k$ 
at any  $k=k_{\rm max}$ 
integrated  over  all  values  lower  than a fixed
value  $P^*_k$  can be expressed as  
(see also, e.g. \citealt{bbks86},  \citealt{fkp94})   
\begin{equation}
{\cal F}(P_k < P^*_k,\  \lambda) = 1 - \exp(-\lambda\ \cdot P^*_k )
\, \, \, \, {\rm at} \, \, \, \,  P^*_k  \geq  0,
\label{calF}
\end{equation}
where $\lambda = \lambda(k)$ 
is a parameter of the exponential distribution
determined by a reciprocal  
mean  (mathematical expectation)
peak  amplitude  M$[P_k] = \langle P_R (k) \rangle$,
i.e. $\lambda (k) = \langle P_R (k) \rangle^{-1}$.
Eq.~(\ref{calF}) allows one  to obtain  significance of a peak  
in 1D  power spectra
when it is possible to calculate
{\it ensemble-averaged}  or  {\it volume-averaged} amplitudes
of spectra  at a given $k$. 

Following Paper~II  we apply this technique
to determine the significance levels of peaks 
in  1D  power spectra 
$P_X(k)$  calculated  using   Eq.~(\ref{PXk}). 
This way  the  significance levels can be found 
from Eq.~(\ref{calF}) 
assuming  that  $\lambda(k) = \langle P_X(k) \rangle^{-1}$,
where $\langle P_X (k) \rangle$ is an
average power spectrum calculated  by rotating of the $X$-axis
within some  {\it background} area  in the sky 
in which no significant peaks  ($ \sim  3 \sigma$)  
have been found
in the  interval of  $k$ we are interested in. 

Eq.~(\ref{calF})  allows us
to calculate  fixed confidence probabilities  
for different  $k$  and  relate  them 
in by one  smooth curve 
to outline  an  appropriate  significance level
over the entire interval under study ($0 \leq k \leq 0.3$).
Such curves  can  be  exploited
as a measure  of significance  for   
spectral   peaks 
obtained   at  any  $k$. 

\section{Rectangle  region.  Radon transform}
\label{sec:rectang}  

\begin{figure*}    
\begin{center}
\includegraphics[width=0.415\textwidth]{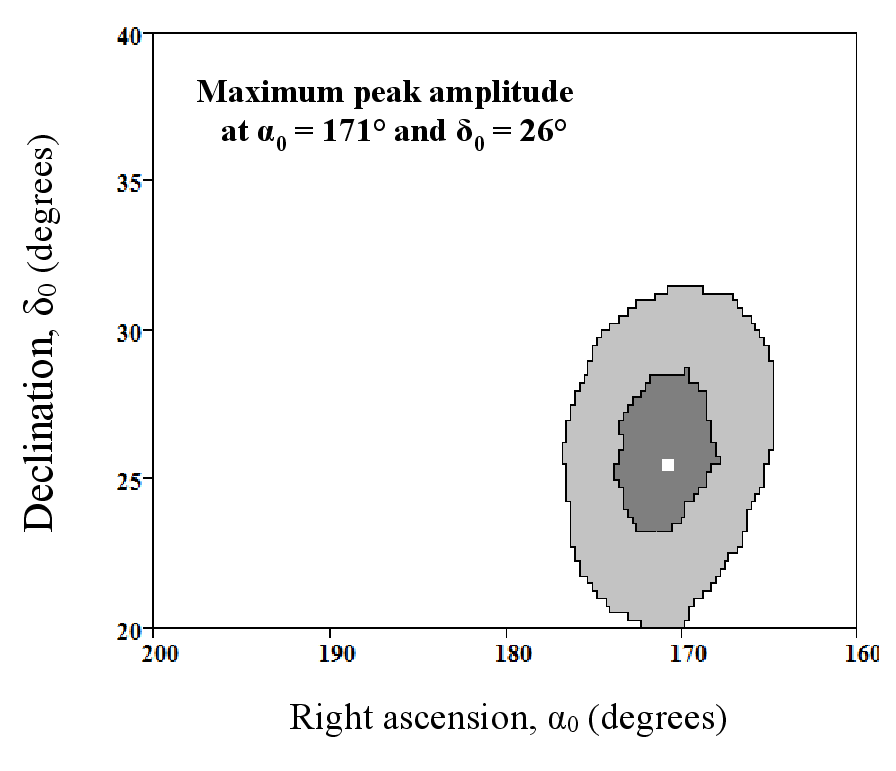}%
\hspace{5mm}
\includegraphics[width=0.450\textwidth]{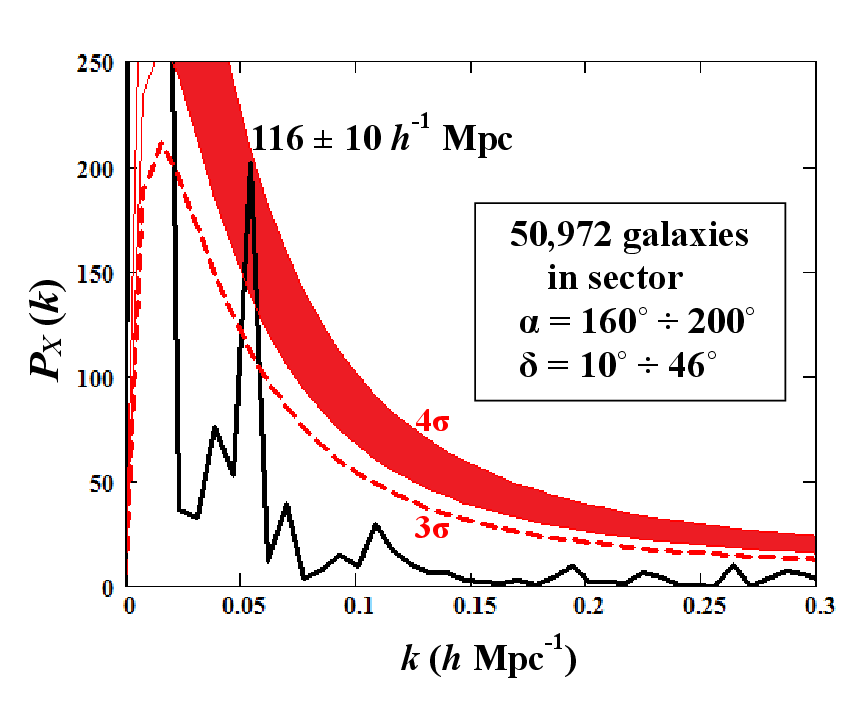}%
\caption{
(Colour online)
{\it Left panel:} 
areas of fixed confidence levels ($\beta = 1-p $)
of the peaks in 
1D power spectra $P_X(k)$  
at  $k_{\rm max} = 0.054~h$~Mpc$^{-1}$
calculated using the Cartesian   
coordinates  of galaxies  projected  
on various axes $X$   
(falling  in a fixed  range   $464 \leq  X  \leq  1274~h^{-1}$~Mpc)
sequentially rotated 
relative to each other by $1^\circ$ 
within the  region  $160^\circ \leq \alpha \leq 200^\circ$
and $20^\circ \leq \delta \leq 40^\circ$
(bounded by dotted lines in Fig.~\protect{\ref{f:rect}}).
The marked  areas  are:
light grey -- $\beta  = 0.998$\ $(3\sigma)$,
grey -- $\beta=0.999936$\ $(4\sigma)$;
white  square indicates  the  direction 
of  the maximum  peak  height  
$X_0$ ($\alpha_0$, \  $\delta_0$).  
{\it  Right panel:} 
1D power spectrum $P_X(k)$ 
({\it solid line}) with a maximum peak (at  $k=k_{\rm max}$) 
calculated for  projections of the 
galaxy  coordinates  on  the $X_0$-axis;  
significance level  $3\sigma$ ({\it dashed line}) 
and   a variety  of  
significance levels   $4\sigma$  within 
the  upward  expanding  band 
with a width equal to the variance
built  for  25 separated samples 
of the {\it background} power spectra 
(see text).   The inset shows the total sample 
of studied  galaxies  and the  $\alpha$,  $\delta$ 
intervals   (light grey  rectangular region  in Fig.~\protect{\ref{f:rect}})  
including  the Equatorial  coordinates of all galaxies 
in this sample;   symbol  `$\div$'  
means an interval, e.g. $\alpha_1 \leq \alpha \leq \alpha_2$.  
}
\label{f:PXk}
\end{center}
\end{figure*}

We  start by orienting  the  $X$-axis along the direction
with coordinates $\alpha=160^\circ$ and $\delta = 20^\circ$
(lower right corner of 
the rectangle region in   
the {\it left panel} of Fig.~\ref{f:PXk}, 
coinciding with 
the smallest rectangle in Fig.~\ref{f:rect}).
We build  the  1D distribution of projections on this axis of
all Cartesian coordinates of galaxies
falling into the entire light grey rectangular area  
in Fig.~\protect{\ref{f:rect}} 
($160^\circ \leq \alpha \leq 200^\circ,\ 
10^\circ  \leq  \beta \leq 46^\circ$) 
and calculate  using Eq.~(\ref{PXk})  the 1D power spectrum 
along  this  direction. 
Then we rotate sequentially  the axis $X'$ 
shifting the right ascension or declination
with a step $1^\circ$ and scanning in this way 
the entire area  shown  on  the  left  
panel of Fig.~\ref{f:PXk}. 
Such  rotations  of  the moving CS require only two
Euler angles, $\alpha_{\rm Eu} = \Delta \alpha$ and 
$\beta_{\rm Eu} = \Delta \delta$, where  $\Delta \alpha$
and  $\Delta \delta$ are respective rotation angles.

In such a  way
one can determine  the existence  of  dominant peaks
in the power spectra 
at $k \sim  (0.05 - 0.07) ~h~{\rm Mpc}^{-1}$
for  a number of directions $X'$  and  delineate 
areas  of increased significance 
of the  peaks. 
Likewise   one can find
an axis  $X_0$ with   the  Equatorial 
coordinates $\alpha_0$  and  $\delta_0$  
along which the 1D distribution 
of the coordinate  projections  shows
the maximum peak height at  indicated $k$.

The results of our calculations 
of the cumulative 1D power spectra  
are represented in Fig.~\ref{f:PXk}. 
The   {\it left  panel} in  Fig.~\ref{f:PXk}  shows two confidence 
areas (two shades of grey) in the sky indicating  peak amplitudes 
(for  a  scale  $2\pi/k_{\rm max} = 116~h^{-1}$~Mpc) exceeding 
the significance levels  $3\sigma$ (light grey) and 
$4\sigma$ (grey), respectively.  
Maximum value of the peak
is achieved  along  the direction of $X_0$ 
with coordinates $\alpha_0=171^\circ$
and $\delta_0=26^\circ$ (small white square).    
Colorless area (``empty''  region) corresponds to
the absence of peaks in 1D spectra,
whose significance reaches  $\sim 3 \sigma$.

The   {\it right  panel}  represents  the  power spectrum  $P_X(k)$ 
calculated for normalized 1D distribution   (Eq.~\ref{NNX})  
of  the   galaxy coordinate projections  
on  the  $X_0$-axis with $\alpha_0$ and $\delta_0$ indicated above. 
The  dashed line  show 
significance levels evaluated 
as it is  described in Sect.~\ref{sec:bd} with
using Eq.~(\ref{calF}).   The  average   
1D  {\it background} power spectra,  
$\langle P(k) \rangle$, 
calculated  by  scanning over  the left  (``empty'')  
half of the entire  {\it left panel} 
in  Fig.~\ref{f:PXk}, i.e at
$180^\circ \leq \alpha \leq 200^\circ$ and  
$20^\circ  \leq \delta \leq 40^\circ$
(total  $21 \times 21 = 441$ power spectra).
In this way one can  obtain 
the  mean  significance  level
($3 \sigma$  and $4 \sigma$  
in  Fig.~\ref{f:PXk}).

However, all power spectra  calculated for close
$X$-axes  are correlated  (not  independent).
Therefore, to  calculate 
the mean  variance of  given significance  
we use a special  procedure of separation of  $X$-directions 
within the {\it  background}  field.  
We  found  that  correlations between 
the power spectra in different  $X$-directions
decrease   to values $<  0.5$
at  angles of $\ga 5^\circ$ between the X-axes.
Thus we select a sample of $16$
independent $X$-directions each of them is separated 
by $5^\circ$  from its nearest neighbors.

Then we build 25 such non-overlapping  samples,  
employing  sequential shifts of the entire grid by $1^\circ$
along $\alpha$ and$/$or $\delta$ axes.    
Using Eq.~(\ref{calF}) and 
pre-computed average power spectra $\langle P(k) \rangle$
we calculate
a fixed  significance level  (e.g.  $4~\sigma$)
for each sample (grid) 
at all considered $k$  from  the  entire   interval  $0 \leq k \leq 0.3$.
Now we can estimate the variance of 25  smoothed dependences on $k$ 
calculated  for the  fixed level of significance
(treating them as random ones for each $k$).
The result of  such  estimations  is  shown  in  the  right  panel 
of  Fig.~\ref{f:PXk}
as  the upward  expanding band  of  the  $4 \sigma$ significance. 

The main  feature  of these calculations 
is  the use of the   3D  Radon  transform  
(e.g. \citealt{deans07}) 
along the selected axis  
forming  the entire sample of  points. 
The resulting 1D distribution
turns out to be sensitive to the presence of  rarefied   quasi-periodic 
components in the spatial distribution of galaxies.
As a result one can see
that the dominant peak   in   $P_X(k)$  at
$k = k_{\rm max}$ 
reaches the upper edge  of  the $4 \sigma$ 
band  and  has  a   height  (amplitude modulus squared) 
about 200 (slightly more).

Worth  to compare  the  result  shown
in Fig.~\ref{f:PXk}  with  results  of Paper~II
represented in Fig.~4 (data of the SDSS DR7)
and  in the right panel of Fig.~7 (black curve;\  
data of the SDSS DR12).   The last two   
indicated  figures refer to
the galaxies observed through 6 selected  narrow 
sectors  belonging to   the  same  large 
rectangular area as outlined by the dashed
lines in Fig.~\ref{f:rect}. 
Note  that  the  scanning area  by the $X$-axis 
was  chosen the same in all  compared   cases, i.e.
$160^\circ \leq  \alpha \leq 200^\circ$   and 
$20^\circ \leq \delta \leq 40^\circ$, 
as well as  the same interval  along $X$
($464 \leq X \leq 1274~h^{-1}$~Mpc)
was  considered,   
to ensure  the same conditions for  the Radon transforms
in all directions under  study.

One can  notice that  significant peaks  
in the power spectra 
were  obtained at the same 
$k = k_{\rm max} = 0.054~h~{\rm Mpc}^{-1}$ 
with  close  directions  of  the  $X_0$  axis 
(maximum  peak):  
$\alpha_0 = 176^\circ,\ \delta_0=24^\circ$
for the DR7 catalog  and 
$\alpha_0 = 175^\circ,\ \delta_0=25^\circ$
for the  DR12  LOWZ  catalog in Paper~II. 
Moreover,  the areas of increased significance 
of the peaks  also  largely  overlap.

It can be seen that the results of this work are well
consistent with the results of  Paper~II, 
although we use more stringent significance criteria 
in the present  study.   Actually,
the significance of the peak 
in the right panel of Fig.~\ref{f:PXk}
does not exceed 
$4 \sigma$, while  its  height   is about 200.
That  is noticeably greater than 
the height of the considered peaks
in   Fig.~7  ( black curve in the   right panel) 
of  Paper~II %
\footnote{Mainly because  
we used extended  interval of $X$ 
for such assessments  in Paper~II
and different areas in the sky (6 sectors) 
for the basic sample of galaxies.}
whose  significance  
has been assessed   as  $\ga 5 \sigma$.

If we do not carry out the gap filling procedure
(described in Sect.~\ref{sec:data})
in the distribution of galaxies included in the LOWZ catalogue, 
then we get the same peak
in the power spectrum  as 
in the {\it right panel} of Fig.~\ref{f:PXk} 
even with a height of 204.6 (instead of 201.3) 
at statistics of 48,776 galaxies (instead of 50,972). 
Thus, the incompleteness of galaxies 
(see Fig.~8 by \citealt{reid16}) 
does not affect the main result.

\section{Rectangle  region.  Rotating cuboid }
\label{sec:cub}  

\begin{figure*}
\begin{center}
\includegraphics[width=0.448\textwidth]{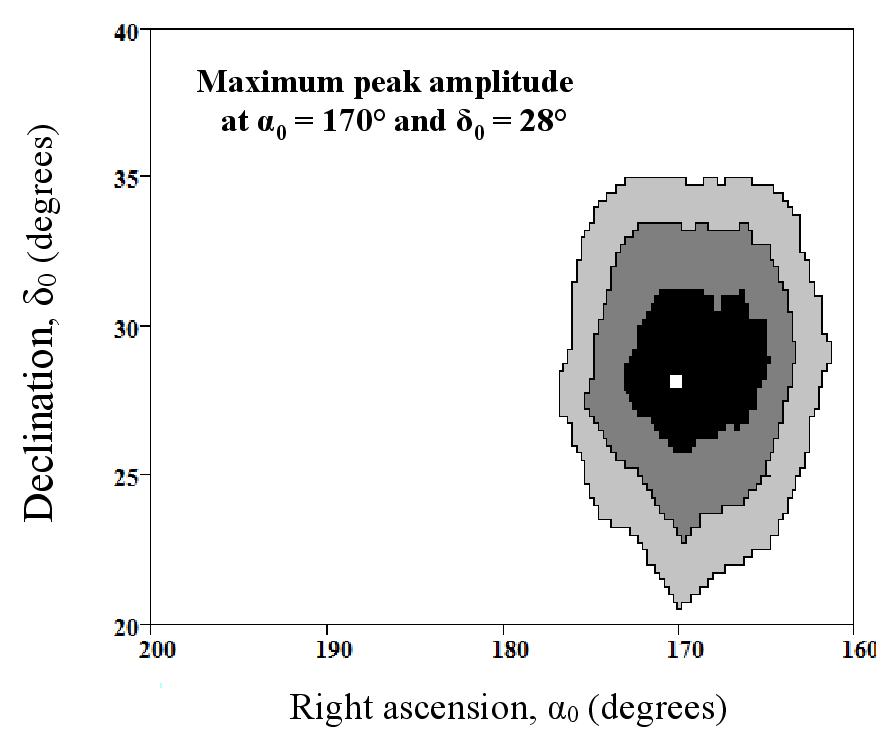}%
\hspace{5mm}
\includegraphics[width=0.442\textwidth]{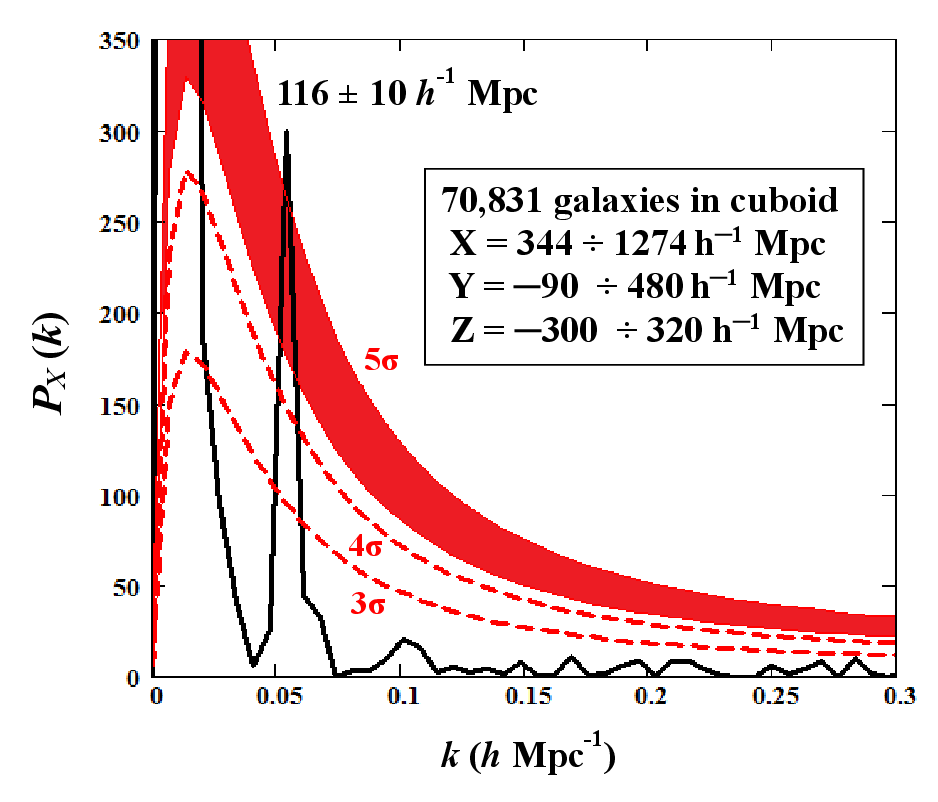}%
\caption{
(Colour online)
Results of similar calculations as in Fig.~\protect{\ref{f:PXk}}
but produced  for the samples  of  Cartesian  
coordinates  of   LOWZ   galaxies  falling into  sequentially rotated cuboid 
with  fixed  boundaries   of   coordinates  $X,\ Y,\ Z$ indicated  
in the inset  on the  ({\it right panel}).
{\it Left panel}:\  
areas of  the  confidence levels  similar to
the  {\it left panel} of Fig.~\protect{\ref{f:PXk}}, 
but with the addition of $5 \sigma$ 
significance  area   ($\beta=0.9999994$,\ {\it dark grey}).
The white square also indicates the direction $X_0$  of the maximum 
spectral peak height ($\alpha_0$ and  $\delta_0$).
{\it Right panel}:\  1D power spectrum   ({\it solid line})
with the maximum peak (at $k = k_{\rm max} = 0.054~h~{\rm Mpc}^{-1}$)   
calculated for the 1D distribution of  coordinate projections on $X_0$-axis;
significance levels  $3\sigma$ and $4\sigma$ ({\it dashed lines}) 
and the  upward  expanding band  with  diversity  of 
significance levels   $5\sigma$ are  calculated  using  power spectra
obtained the same way   
as in  Fig.~\protect{\ref{f:PXk}}. 
The total number of galaxies  falling into  the cuboid
oriented in the $X_0$ direction is also indicated in the inset.
}
\label{f:cuboid}
\end{center}
\end{figure*}

Fig.~\ref{f:cuboid} demonstrates  results 
of applying another method  of  searching  
for  the traces  of  periodicity in the same scan area 
of the $X$-axis, as in  the  {\it  left panel} of Fig.~\ref{f:PXk}.
For this purpose  we build a  {\it cuboid}  (rectangular parallelepiped) 
in space  with fixed  the  Cartesian coordinates $X, Y, Z$
(see below),   the $X$-axis  being responsible 
for the orientation of the cuboid in space.  
We  define   vertices and  faces  of the cuboid 
and calculate the projections of the coordinates of all galaxies
from the  LOWZ catalog,  falling into  the cuboid,  
onto the direction of  $X$-axis.
This  direction is determined by the angles  $\alpha$ and $\delta$ 
in  the Equatorial  CS.  As in  the  Sec.~\ref{sec:rectang}, 
we build  the 1D  distribution of these projections and calculate its 
power spectrum,  then we rotate the $X$-axis along with the cuboid 
around the origin  $X=0$  and repeat the whole procedure  at  each 
step  of  the cuboid's rotation.

As the first step, the $X$-axis is aligned 
with the direction of the maximum amplitude 
of the peak in the power spectra  
shown in the {\it left panel} of  Fig.~\ref{f:PXk}, i.e.    
$\alpha_0=171^\circ$ and $\delta_0=26^\circ$.   
Then we move the vertices of the cuboid, 
changing its orientation but 
without changing  its shape and size, 
until the peak in the power spectrum 
within the interval $ 0.05 < k  < 0.07$
reaches the  maximum.  The coordinates of the vertices and 
positions of the edges   found in this way are 
indicated in the inset in the {\it right panel} of Fig.~\ref{f:cuboid}. 
Thus, the positions of the faces closest to the observer
and the most distant from him  
are  defined  by   a  condition 
$344 \leq X \leq 1274~h^{-1}$~Mpc. 
Let us  notice  
asymmetric shifts  of the vertices
along the $Y$- and $Z$-axes relative to the $X$-axis
($-90 \leq Y \leq 480~h^{-1}~{\rm Mpc}$,\  
$-300 \leq Z \leq 320~h^{-1}~{\rm Mpc}$).

The vertex coordinates in the  Cartesian CS
obtained in  this  way  
rotating together  with the cuboid  do not change
in the process of scanning by  the $X$-axis (with a step $1^\circ$)
of the entire area  $160^\circ \leq \alpha \leq 200^\circ$
and $20^\circ \leq \delta \leq 40^\circ$.
For all valid X-axis orientations
and subsequent constructions of cuboids with respect to these directions, 
the boundaries of the volume of cuboids
capture all detected galaxies inside the large rectangle in Fig.~\ref{f:rect}.
This  expands the statistics of galaxies compared  to  Sect.~\ref{sec:rectang}. 
On the other hand,  almost all positions of the cuboid 
(with  minor exception) 
locate  within the  large rectangle. 

The {\it left panel} of Fig.~\ref{f:cuboid} looks
similar to the left panel of Fig.~\ref{f:PXk}.
Actually,  a highlighted  area  inside the 
scanning  sector of the $X$-axis  
is quite consistent with the  oval-like  area 
in the left panel of Fig. \ref{f:PXk}.  
Both  areas  correspond  to   the 
increased  significance  of  peaks 
in the power spectra  with that  difference 
that the {\it left panel} of Fig.~\ref{f:cuboid}
contains a (dark grey) region  with  significance  $\ga 5 \sigma$, 
which is missing in Fig.~\ref{f:PXk}. 
The small white square also indicate the  direction  of the maximum   
$X_0\quad  (\alpha_0 = 170^\circ, \  \delta_0 = 28^\circ)$   
at $k=k_{\rm max}=0.054~h$~Mpc$^{-1}$. 
This direction is quite close to the   
direction of the  maximum  spectral  peak  obtained  for
the Radon transforms (Fig.~\ref{f:PXk}).

The {\it right panel}  of  Fig.~\ref{f:cuboid} is  also
organized similar  to  the right  panel   in  Fig.~\ref{f:PXk}.
A very high peak  in the 1D power spectra 
(calculated for the $X_0$-direction)  
with  a  height  $\sim 300$
at $k=0.054~h~{\rm Mpc}^{-1}$  is visible. 
Two significance levels 
($3 \sigma$ and $4 \sigma$)  
represented  by  dashed lines
and the wide band  corresponding to  $5 \sigma$  
are also calculated  in the same way as
it   is  described in Sec.~\ref{sec:rectang}
for $4~\sigma$ level. 

The main difference between the two methods is that 
during  rotations  of the cuboid 
there is a partial exchange of galaxies from the
general  array 
captured by the cuboid in   different  positions,
while in  the approach of  Radon transforms  we use
projections onto different $X$-axes 
(but with fixed boundaries of the interval 
$X_1 \leq X \leq X_2$)
from a single  sample of
galaxies  bounded by a smaller rectangle
in Fig.~\ref{f:rect}.  It can be concluded 
that scanning of the area
shown in the {\it left panel} of Fig.~\ref{f:cuboid}
by  the  $X$-axis,
accumulating   all  projections  of  galaxy  coordinates
falling into the cuboid, gives a more efficient  way to
search for quasi-periodicity in the space distribution
of the  galaxies.  Moreover, 
it can be assumed that projections of
coordinates of galaxies falling into the cuboid 
onto  the  $X_0$-axis
make it possible to roughly localize 
the spatial regular structure  if  it  is  really
exists.%
\footnote{Note  that the Radon transform allows one only to
determine  the main direction  $X$ of quasi-periodicity,
but not its localization in space.}

Note that comparison with results 
obtained in a similar way for the unfilled gaps 
in the LOWZ catalogue,  
occurring within the large rectangle in Fig.~\ref{f:rect}, 
leads to the same conclusion as the one 
at the end of Sect~\ref{sec:rectang}.
Indeed, we  get the same peak
in the power spectrum, as in the 
{\it right panel} of Fig.~\ref{f:cuboid}, 
whose height is 297 (instead of 299) 
with statistics of 67,359 (instead of 70,831).

\section{Homogeneous sample modeling}
\label{sec:mod}  

\begin{figure*}  
\includegraphics[width=0.436\textwidth]{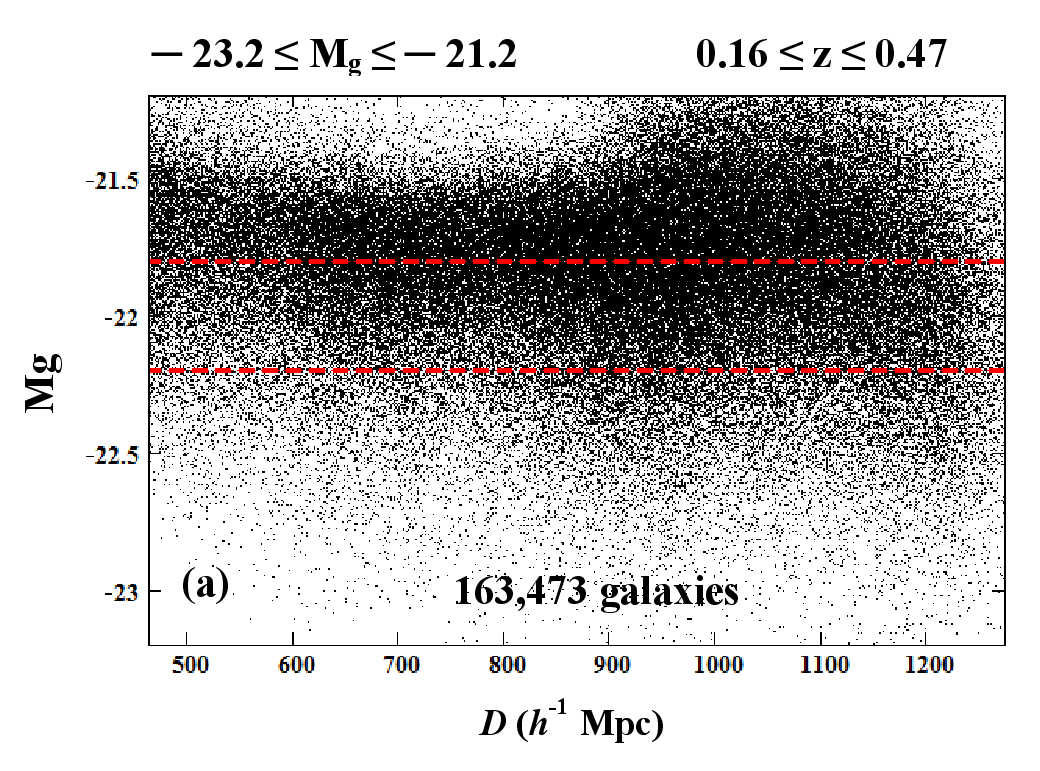}%
\hspace{5mm}
\includegraphics[width=0.462\textwidth]{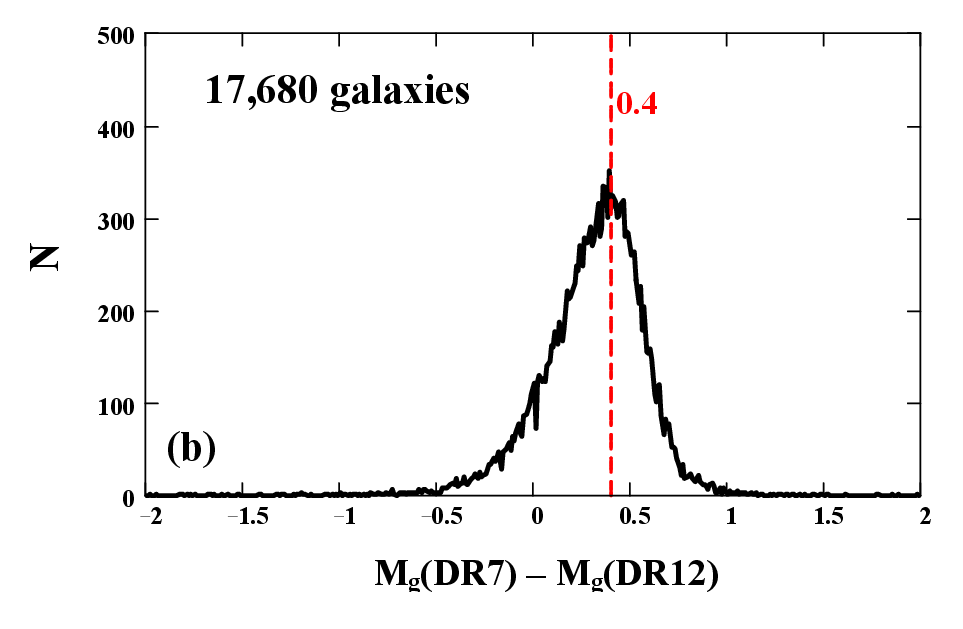}%
\vspace{5mm}
\includegraphics[width=0.460\textwidth]{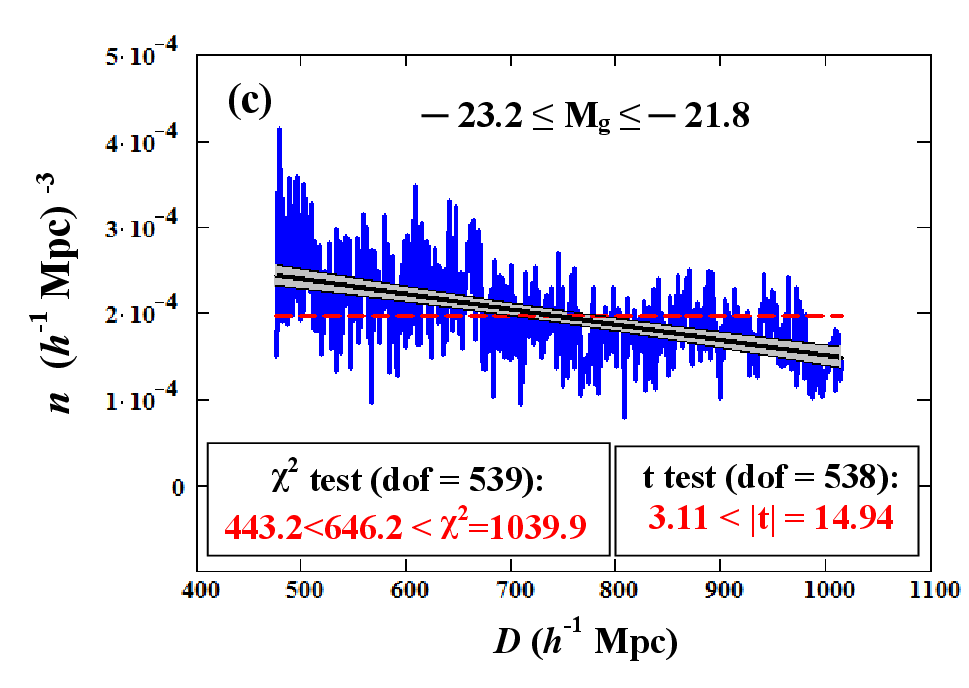}%
\hspace{5mm}
\includegraphics[width=0.474\textwidth]{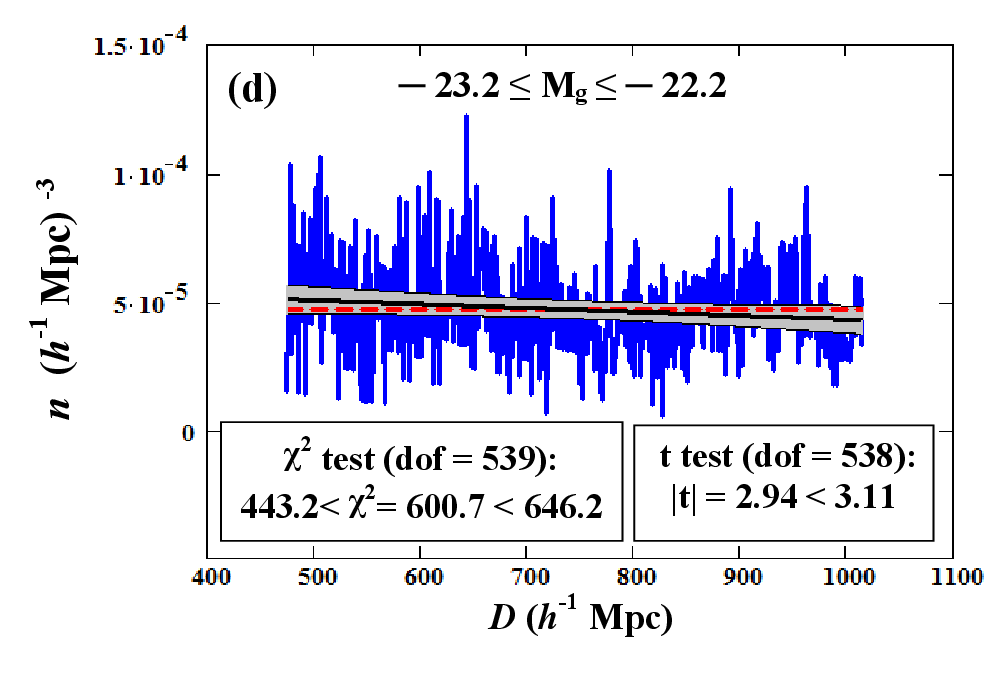}%
\caption{
(Colour online)
{\it  Panel}~(a):   distribution of the LOWZ galaxies, 
observed within the angular rectangle
$140^\circ \leq \alpha \leq 230^\circ$  and  
$0^\circ \leq \delta \leq 60^\circ$, 
and the interval of redshifts  $0.16 \leq z \leq 0.47$,
on the  $\left( D(z),  {\rm M}_g \right)$-plane, where
$D(z)$ is  the  comoving distance and ${\rm  M}_g $ is   
 the absolute  $g$-band magnitude.
Two  horizontal dashed  lines  indicate  the levels 
${\rm M}_g = - 21.8$ and $-22.2$ from top to bottom,
respectively. The total number of galaxies 
included in the distribution 
is indicated at the bottom. 
{\it Panel}~(b):\  distribution of  
differences of absolute magnitudes 
${\rm M}_g $~(DR7)  -  ${\rm M}_g $~(DR12) 
(on condition $ -23.2 \leq {\rm M}_g  \leq - 21.8$ ) 
of  the  same galaxies 
registered in  two  different   catalogues:   DR7\     and    DR12\  LOWZ
(see text).  
The intervals of angular coordinates $\alpha$ and $\delta$,
and  redshifts $z$ are  the same as in  the {\it panel}~(a).
The vertical dashed line 
indicates  the maximum of the distribution. 
The number of galaxies included in the distribution 
is also shown. 
{\it  Panel}~(c):\  average  number density  $n(D)$
of the LOWZ galaxies  as a function
of the comoving distance 
on the same condition   ($-23.2 \leq {\rm M}_g \leq -21.8$)
as in the {\it panel}~(b)  and   within 
the same angular rectangle region  
as in the {\it panels}~(a) and  (b),\  
but  for a shorter interval  
$464 \leq  D  \leq 1000~h^{-1}$~Mpc \
($0.16 \leq z \leq 0.36$).  
The thick solid  line corresponds to  the linear regression
with the most probable parameters (see text).  
The confidence region ($\beta = 1-p = 0.998$), i.e.  grey strip   
about the solid  slant line,  
is  also shown. 
The horizontal dashed line
corresponds to  the model  $n(D)=const$. 
Results of $\chi^2$- (bottom  left) and  $t$- two-sided  tests (bottom  right)
applied  to  assess a compatibility of the 
horizontal dashed  line with the data   are
indicated in the  insets. 
{\it Panel}~(d):\  Same as  in  the  {\it panel}~(c) but for 
brighter LOWZ galaxies ($-23.2 \leq {\rm M}_g  \leq -22.2$).
}
\label{f:VLS}
\end{figure*}

Fig.~\ref{f:VLS}~(a)   shows the absolute  $g$-band magnitudes ${\rm M}_g$
of the sample of galaxies observed within the  largest  angular  rectangle 
($140^\circ \leq \alpha \leq 230^\circ,\  0^\circ \leq \delta \leq 60^\circ$) 
in Fig.~\ref{f:rect} in dependence on  the  comoving distance $D(z)$ (Eq.~(\ref{D})).
The rest-frame  absolute magnitude of the $i$-th galaxy with the redshift $z_i$
is defined according to \citet{beck16} as  
${\rm M}_{g, i} = m_{g, i} - 25 - 5 \log{\left[(1+z_i) D(z_i) / (h^{-1} {\rm Mpc})\right]} - K_{g, i} (z=0)$,
where $m_{g, i}$ is  the apparent  $g$-band magnitude,\ 
$K_{g, i} (z=0)$ is the corresponding $K$-correction.
 
The considered interval $464 \leq D \leq 1274~h^{-1}$~Mpc  
corresponds to   redshifts  $0.16 \leq z \leq 0.47$.
Each point on the  $\left(D,\  {\rm M}_g \right)$-plane represents
a galaxy from the   DR12\ LOWZ catalog.
In this section, we neglect  the  existence of the area 
of incompleteness   in the LOWZ  sample
({\it chunks} $3 - 6$ in  the  notation  of   \citealt{reid16})  
 and do not use the gap-filling procedure  described in Section 2. 
The results obtained below  
indicate again that the region of incompleteness 
does not  principally affect  the  appearing
of   quasi-regularity traces.
The total  number of points  (sample size) 
within the interval $-23.2 \leq {\rm M}_g \leq -21.2$ 
is 163,473 galaxies. 
 
The absolute $g$-band magnitude ${\rm M}_g$ 
were  found  in  the  database  SDSS  SkyServer
(e.g. \citealt{beck16})%
\footnote{http://skyserver.sdss.org/dr12/en/help/docs/realquery.aspx}
\footnote{The galaxy (photometric) identifier 
(19-bit number) p.ObjID\   queried from the basic  catalogue 
of galaxies,  {\it photoprimary p}, allows   one  to  join  
the requested  data on  $\alpha, \beta, z_s$  
in the spectroscopic  table,  {\it specphotoall s},     
with  the  data  on  ${\rm M}_g $ in the  photometric table   
{\it photoz ph}.}  
using  additionally 
the selection conditions  given in the formulas 
(9 - 12) by  \citet{reid16},  as well as  keeping 
the  fixed region of equatorial   
coordinates  and  the  fixed interval  $z$  (indicated above).
Note, however, that
the error of such an identification  remains rather uncertain.

In order to determine ${\rm M}_g$ of analyzed galaxies  
in the LOWZ catalogue
we have created a new table based on 
the SDSS SkyServer database containing 
three galaxy identification parameters:
two Equatorial coordinates $\alpha$,\  $\delta$ and 
spectroscopic redshifts  $z_s$ 
and their photometric g-band absolute magnitudes M$_g$. 
The first three parameters were used to search for 
the same galaxies in the 
DR12 LOWZ  catalogue  ($galaxy$\_$DR12v5$\_$LOWZ$\_$North.fits.gz$) 
with the subsequent assignment to them  of 
the corresponding M$_g$. 
As a result 163,473 galaxies included in both the lists were selected,
and    their  absolute  magnitudes  ${\rm M}_g$  were  found. 

In Fig.~\ref{f:VLS}~(a)
one  can see a inhomogeneous  distribution of
points with comoving distance $D$, in particular,  for brighter
objects below the ${\rm M}_g < - 21.8 $  
(upper dashed line).  However, for even  brighter 
${\rm M}_g < - 22.2$ (lower dashed line) 
but rarer galaxies, the visible  distribution becomes 
more   homogeneous. 

Fig.~\ref{f:VLS}~(b) demonstrates  a systematic shift of the $g$-band
absolute magnitudes $-23.2 < {\rm M}_g < -21.8$  
between the data of two catalogues of the LRGs
DR7\quad   
[Full\  DR7\ LRG\  sample presented  by   \citet{kazetal10}]%
\footnote{https://cosmo.nyu.edu/$\sim$eak306/SDSS-LRG.html} 
and DR12\  LOWZ\  catalogue  in combination with  the database
SDSS  SkyServer  as described  above.  
The figure represents the distribution of the difference
between absolute magnitudes of two catalogues related 
to a sample of the same LRGs,  
the  intervals of the equatorial 
coordinates  ($\alpha$ and $\delta$)
and  the comoving distances  
($464 \leq  D  \leq 1274~h^{-1}$~Mpc)  
being   the same as  in the {\it panel}~(a). 
Under these conditions, the total number of  LRGs 
common for these two  catalogues  is  17,680. 

The  shift  $\Delta {\rm M}_g \approx 0.4$ 
represented  in  Fig.~\ref{f:VLS}~(b) 
makes it  difficult  a direct  comparison  of
galaxy  distributions falling within the same  intervals 
of  ${\rm M}_g$   from the DR7  and   the 
DR12 LOWZ catalogues.   
In particular, the conditions of a volume-limited sample 
($n(D) \approx $~const  or  $n(z) \approx $~const) 
are satisfied  for  the  subsample   DR7-Bright   presented  by  \citet{kazetal10}
(at  $-23.3 \leq {\rm M}_g \leq -21.8$ and $0.16 \leq z \leq 0.44$), 
but this does not match  with the properties 
of the sample obtained in the present  work.

Figs.~\ref{f:VLS}~(c) and (d) illustrate this  difference. 
They  show
dependences on the comoving distances $D(z)$
of the LOWZ galaxy number densities  $n(D)$,
averaged over  the angular coordinates  
within  the  sky region  indicated above.
Here we use  a bin  $\delta D = 1~h^{-1}$~Mpc
and  a smaller   interval  of  $D(z)$  than in  Fig.~\ref{f:VLS}~(a),    
$464 \leq   D  \leq 1000~h^{-1}$~Mpc  (or  $0.16 \leq z \leq 0.36$).
The {\it panel}~(c)  corresponds to the absolute magnitudes 
$-23.2 < {\rm M}_g < -21.8$\   (17,086~galaxies), while the {\it panel}~(d) --
to  the brighter and  rarer galaxies at 
$-23.2 < {\rm M}_g < -22.2$\  (4251~galaxies).
The small bin size turned out to be more suitable 
to  perform   linear  regression  (see below), 
while the   decrease in the upper limit  of the interval  $z$   provides  almost
constant average number density 
(cf. with  Figure 2 of  \citealt{kazetal10}).

The thick  solid  lines in both the {\it panels} (c) and (d) 
correspond  to the linear 
regression determined by the maximum likelihood method
(e.g.  \citealt{ll84}) \quad 
$n(D)=c_1 +  c_2 (D - \overline{D})$
with the most probable parameters 
$c_1 = 2.0 \times 10^{-4}~(h^{-1}~{\rm Mpc})^{-3}$ and 
$c_2= - 1.8 \times 10^{-7}~(h^{-1}~{\rm Mpc})^{-4}$
for the  {\it panel}~(c),   and 
$c_1 = 4.7 \times 10^{-5}~ (h^{-1}~{\rm Mpc})^{-3}$ and  
$c_2 = - 1.5 \times 10^{-8}~ (h^{-1}~{\rm Mpc})^{-4}$
for the  {\it panel}~(d);  in both cases $\overline{D} = 745~h^{-1}$~Mpc.

The horizontal dashed  lines correspond to the conception  
of  a  volume-limited sample with nearly constant number density.
Results of  $\chi^2$ and $t$ - tests  applied   
to  assess a  compatibility of
these horizontal lines   with  the  data on $n(D)$
are shown in the insets at the bottom  
of  both the panels
(left  and  right,   respectively). 
It can be seen that in the {\it panel}~(c) 
the obtained values of  $\chi^2$  and  $t$ 
located far outside the confidence  region at  
$1 - p  = 0.998$.

On the other hand, the best fit 
in the {\it panel} (c)  at  ${\rm M}_g < - 21.8$ 
gives the  slant  linear  thick  line  that  differs significantly from 
the  horizontal  dashed  one.  
The confidence region  at  $1 - p = 0.998$  about  the  linear  
dependence $n(D)$
is also shown by narrow grey strip restricted by two dashed lines. 
Thus  the hypothesis that $n(D) = const$   
at  ${\rm M}_g < - 21.8$ 
does not statistically consistent with 
the set of $n(D)$ values.

While,  in the {\it panel}~(d)  at    ${\rm M}_g  < -22.2$
the  solid thick line is in a good agreement
with the horizontal dashed line, i.e. the number density
of  relatively  brighter galaxies 
is  close to homogeneous, which corresponds to the 
volume-limited sample.  
This means  that  the close-to-homogeneous    
distribution $n(z)$  presented at  ${\rm M}_g < -21.8$
in Fig.~(2) by \citet{kazetal10} is  likely 
shifted  for  the LOWZ  galaxies   on  absolute magnitude
$\Delta {\rm M}_g  \approx 0.4$ 
in accordance with  Fig~\ref{f:VLS}~(b), although
such  a  shift occurs with a strong loss 
of statistics in the LOWZ  catalog.

\begin{figure*}  
\includegraphics[width=0.448\textwidth]{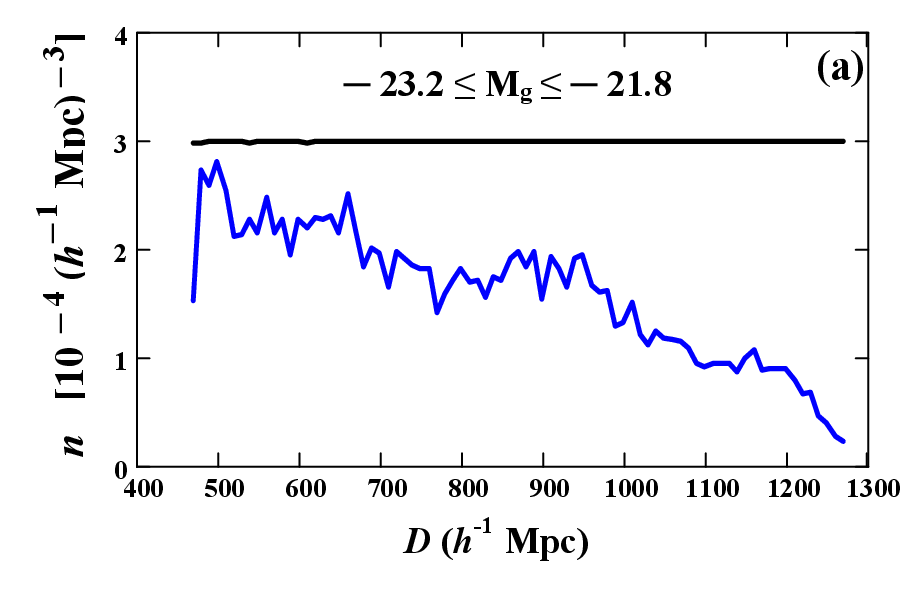}%
\hspace{5mm}
\includegraphics[width=0.448\textwidth]{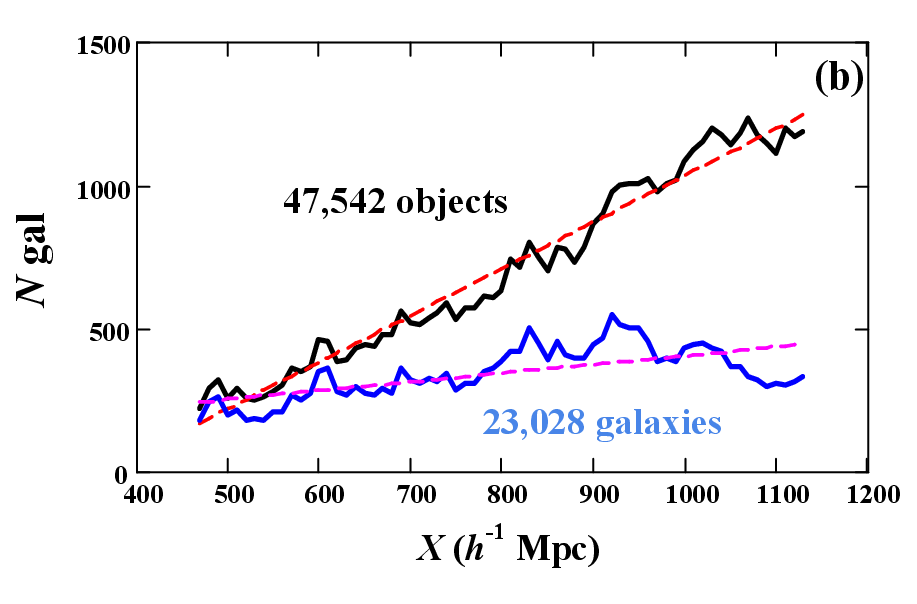}%
\vspace{5mm}
\includegraphics[width=0.476\textwidth]{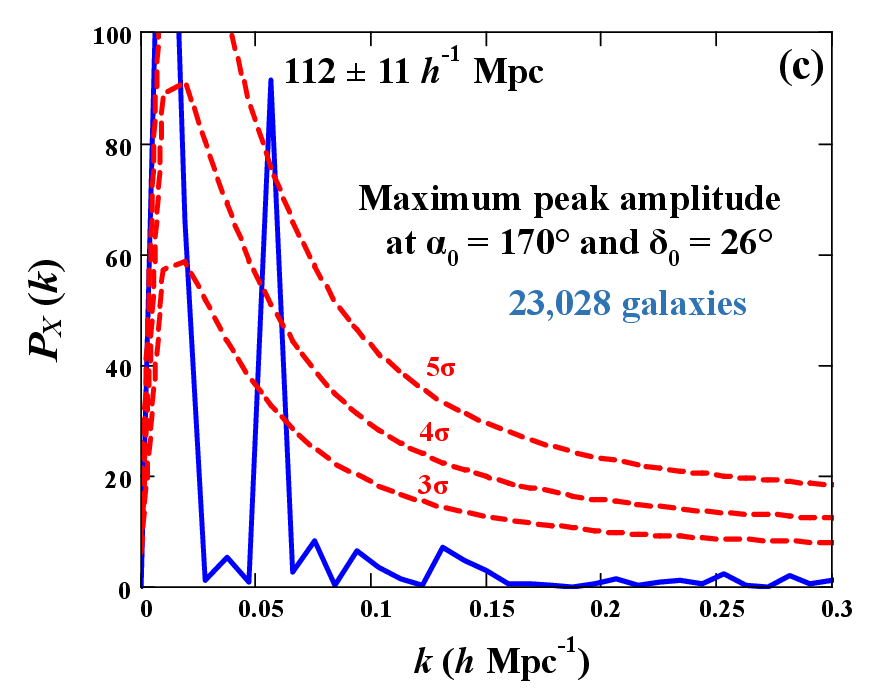}%
\hspace{5mm}
\includegraphics[width=0.462\textwidth]{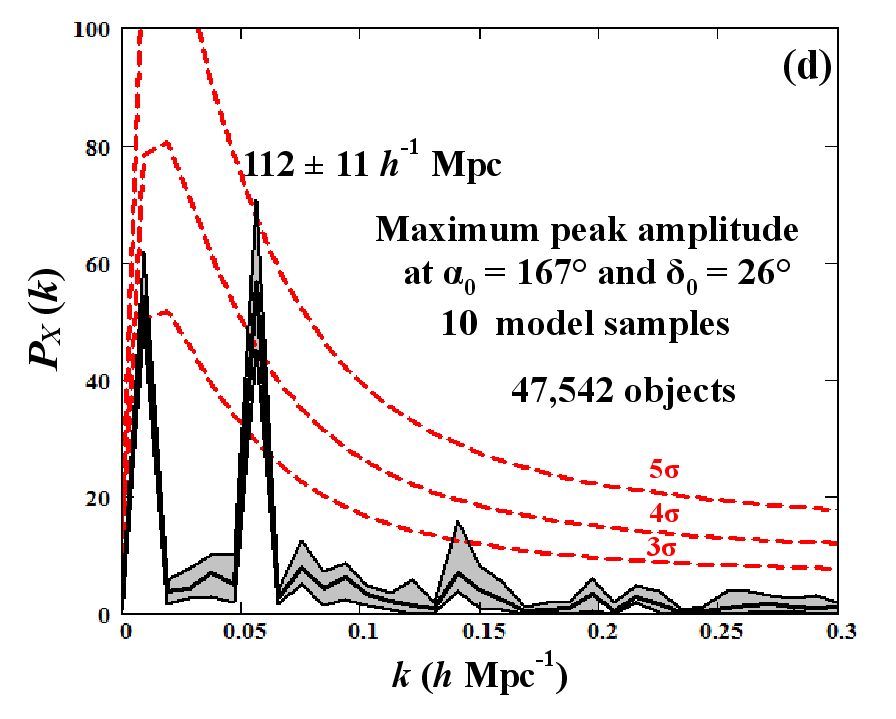}%
\caption{
(Colour online)
The results of modeling of  the reduction 
of possible selection effects  inherent to  the 
statistics of  LOWZ\  DR12 galaxies  at  
$-23.2 \leq {\rm M}_g  \leq -21.8$ 
and  angular  coordinates  located within a smaller (light grey) rectangle
($160^\circ \leq \alpha \leq 200^\circ$ and  
$10^\circ \leq \delta \leq 46^\circ$) 
in Fig.~\protect{\ref{f:rect}}  (see text).
{\it Panel}~(a):\   the  lower   curve   represents
the   number  density  of  galaxies
$n(D)$ in units  $10^{-4}~ h^3 {\rm Mpc}^{-3}$
as  a  function of  the  comoving  distance
$D$ within the full interval  under consideration 
$464 \leq D(z) \leq 1274~h^{-1}$~Mpc;  
upper horizontal line corresponds to the
created sample  of both the real  galaxies and 
artificially added  points 
randomly distributed in comoving  space. 
{\it Panel}~(b):\   
shows two distributions of projections 
of Cartesian  coordinates 
on the  $X_0$-axes   within
the fixed  interval ($464 \leq X  \leq 1134~h^{-1}{\rm Mpc}$); 
two $X_0$-axes are oriented in such a way  
that the dominant peak  height
of  the power spectra 
at  $k = k_{\rm max}=0.056~h~{\rm Mpc}^{-1}$ 
calculated separately for  the sample of  $23,028$  galaxies 
and  the  model   sample   of  $47,542$  objects 
would be maximum;
two directions of  $X_0$  (i.e. $\alpha_0\  and\   \delta_0$) are indicated 
in  the  {\it panels}  (c)  and  (d),  respectively.
The lower  curve   corresponds 
to the sample of  galaxies,   while  the  upper  rising  
curve   is  an example of  model   sample;
dashed lines depict linear trends of  the  two distributions.  
{\it Panel}~(c):\   shows the 1D power spectrum calculated for 
the   projections  of  Cartesian coordinates 
of  real   galaxies  on the  $X_0$-axis.
{\it Panel}~(d):\   plots   the  1D power 
spectrum (thick black line)  averaged over 10 power spectra 
calculated  for  randomly selected  model samples  
and  a  variations  of  peak amplitudes
(at $0 \leq k \leq 0.3$) 
corresponding to  the same  
10 model power  spectra 
(grey strip bounded by two thin lines). 
Dashed lines  in   {\it panels}~(c)  and  (d) 
indicate significance levels  calculated  in  the  same  manner
as   the dashed lines  in  the {\it  right  panels} 
of  Figs.~\protect{\ref{f:PXk}} and \protect{\ref{f:cuboid}} 
(see text).  
}
\label{f:model}
\end{figure*}

Fig.~\ref{f:model} shows  the  results 
of our model calculations  reducing  
an influence  of possible  selection effects
onto  the inhomogeneous distribution  $n(D)$ of  the  DR12 LOWZ   galaxies.
For this  purpose,  many  objects  (points)  are artificially added to each 
independent  bin  ($\Delta_D = 10~h^{-1}$~Mpc)   of comoving distances
to get a uniform distribution at the level of
the maximum value of  $n(D)$ in the  considered range $D$.
Objects (points) are added to each  bin  $\Delta_D$  randomly, 
so that the two angular coordinates  $\alpha$
and  $\delta$    as well as the comoving distance itself, 
correspond to  uniform  distribution  (Poisson's  distribution in space). 
Joint distribution of these added  objects
and real   LOWZ  galaxies forms a model sample,
represented by the horizontal line in the {\it panel} (a).

The properties of the spatial distribution corresponding 
to the total  model sample can be examined using the procedure 
outlined in Sect.~\ref{sec:bd}  but with some variations. 
To do this, as in  
Sects.~\ref{sec:rectang}  and  ~\ref{sec:cub}, 
a Cartesian coordinate system is introduced, 
which rotates in such a way that the $X$-axis scans 
with a step of $1^\circ$ over the entire area outlined 
by the dotted  line in Fig.~\ref{f:rect},  
i.e. the same area as it is used
in Figs.~\ref{f:PXk} and  \ref{f:cuboid}.    
For each  $X$-direction, projections  of the coordinates 
of both  artificial and real galaxies  (model sample) 
onto this axis  are  produced,
then  1D normalized distribution and its power spectrum are
calculated.  For comparison
all subsequent  calculations are  
also carried out separately   for real  LOWZ  galaxies 
within the same range of values 
${\rm M}_g$,\  $\alpha$,\ $\delta$ and $X$.

It should be noted that for the model sample 
it is not possible to coherently
construct a normalized 1D 
distribution function using Eq.~(\ref{NNX})
and calculate the power spectrum  using Eq.~(\ref{PXk}) 
for  the entire interval  $464 \leq X \leq 1274~h^{-1}$~Mpc. 
This is a consequence of  a complex trend of the dependence $N_X (X)$
having a rather high maximum at  $X \sim 1100~h^{-1}$~Mpc 
and a sharp subsequent decline (not shown in Fig. \ref{f:model}~(b)).
Therefore, we have to  decrease  the interval $X$ 
up to $464 \leq X \leq 1134~h^{-1}$~Mpc,
so that the distribution of the model sample
well described  
(using the least squares method) 
by the simplest linear trend,
which is shown in Fig.~\ref{f:model}~(b). 
Replacing in  Eq.~(\ref{NNX}) the mean value  $S_X$ 
by  a linear trend   $N_L (X) = s_1 + s_2 X$, 
where  $s_1$  and  $s_2$  are  determined  constants,
and  using formula  Eq.~(\ref{PXk}), 
we obtain the power  spectra  for both
the  sample of  LOWZ galaxies  Fig.~\ref{f:model}~(c) 
and   the  model  sample  Fig.~\ref{f:model}~(d).

Fig.~\ref{f:model}~(c)  represents
the power spectrum of  1D distribution $NN(X)$ of  real  galaxies 
with limited  absolute magnitudes  ($-23.2 \leq {\rm M}_g  \leq - 21.8$)
in the direction $X_0$  of the maximum peak.
The spectrum plotted using the linear trend [shown 
at the bottom of Fig. \ref{f:model}~(b)]. 
We  see  that the significance of this  peak 
exceeds  $5 \sigma$.

Fig.~\ref{f:model}~(d) demonstrates  the results 
of power spectra calculations 
carried out for 10 model samples 
using  the same algorithm as described above.
Various realizations of power spectra form a    strip
of variations  (light grey) visible in the panel. 
This  strip, at  $k=k_{\rm max}=0.056$ 
corresponds to the scatter of peaks whose amplitudes lie
in the interval of significance from $4 \sigma$ to $5 \sigma$. 
There are also shown the 
power spectrum averaged over all 10 realizations.
The $X_0$-direction  indicated in the panel (d) 
corresponds to  the highest peak amplitude
among all 10 realizations.   Note that
Fig.~\ref{f:model}~(b)  plots the $N_X(X)$  distribution
for the same direction.

Significance  levels   in Figs.~\ref{f:model}~(c) and (d)
were carried out according to the same  procedure
as described above  for  the significance  levels 
in  Figs.~\ref{f:PXk} and  \ref{f:cuboid},  i.e.  
using the same area  of $X$-directions 
($180^\circ \leq \alpha \leq 200^\circ$ and   
$20^\circ \leq  \delta \leq 40^\circ$), 
in which there are no significant peaks in the power spectra.
Note also a slight  shift   
of  peak centers
in both the power spectra 
relative the position of peaks in
Figs.~\ref{f:PXk} and  \ref{f:cuboid}
but quite consistent with them.
The  shift  possibly related 
to the use of the  linear trend  $N_L(X)$
instead of the mean value $S_X$ in the modification
of   Eq.~(\ref{NNX}).

Thus, the model calculations carried out here  
show  that the artificial intrusion  of random objects, 
which ensures the  homogeneous number density  
$n(z)$ of the model sample, 
does not smooth out the quasi-periodic feature 
in the spatial distribution of galaxies. 
In other words, we are dealing with a rather robust 
quasi-periodicity.

\section{Conclusions and  discussion}
\label{sec:cd}
In this  work
on the base of   SDSS\  DR12\  LOWZ   catalogue
we   study   possible large-scale  quasi-regular structure
in the spatial distribution of cosmologically  distant  galaxies 
at   redshifts  $0.16 \leq   z    \leq 0.47 $,  
or comoving  distances  $464 \leq D \leq 1274~h^{-1}$~Mpc.

Unlike our previous  papers 
(e.g.\  \citealt{rk14},\  Paper~I,\   Paper~II ), 
here we do not consider the radial distributions 
of  objects in space  but use others,  as it  turned 
out to be  more sensitive,   methods of  registration of   
possible quasi-regular 
structures located at cosmological distances. 
We apply the method proposed in Paper~II 
of projecting the Cartesian coordinates of galaxies 
onto different $X$-axes 
sequentially rotated 
within definite region on the sky.

We use two modifications   of   projections
in each rotational state of the $X$-axis,
namely, (i) the projections onto the $X$-axes  of the Cartesian coordinates 
of all galaxies  located  in the definite  interval $z_1 \leq z \leq z_2 $ and 
detected  within   a certain region in the sky 
(light grey rectangle in Fig.~\ref{f:rect}),
(ii) similar projections on the $X$-axis  of  
coordinates of all galaxies  falling
into  the  imaginary   rotating   {\it cuboid} whose faces (edges and vertices)  
are  defined in the  $XYZ$-coordinate system  
rigidly connected with  the cuboid.
When  rotating,   the  cuboid  captures  unlike the case (i) 
different  samples of  the  LOWZ  galaxies  
located mainly   within  the  large   rectangle   
in Fig.~\ref{f:rect}.  In  each rotational  state 
we calculate  1D distribution 
of the projections  of galaxy coordinates  
(captured in this way)  
onto  the  $X$ axis
with  a  fixed   interval   
$X_1 \leq X \leq X_2$.

In both the cases  
we plot the  normalized  1D  distributions along each $X$-axis and calculate the 
corresponding power spectrum. 
Thus, the direction of the maximum peak  heights  
$X_0$\  ($\alpha_0,\ \delta_0$) of  the power spectrum
in a narrow range   $0.05 < k  < 0.07$ 
can  be   found   among a  set  of   directions $X$
with various  power spectra.
In the case (i) we 
perform the  so-called  discrete 3D Radon transform
along a given  $X$-axis  and calculate its power spectrum,   
keeping in mind that the  real    direction   of  quasi-periodicity 
can  be  parallel to the found axis  $X_0$ 
but arbitrary  shifted in space.
While in the case (ii) we have a chance  at least roughly to  localize 
this structure in space.

To  make  assessments of  significance   
we use  
an approach of  exponential probability [see Eq.~(\ref{calF})] 
to get  a value of randomly distributed   
peak  heights in the  power spectra.
The key  of  such assessment   is
calculation of the average
power spectrum  $\langle P(k) \rangle$ 
with  variable  variance
throughout the whole  region,  $0.0 < k \leq  0.3$,   under consideration.
These spectra were averaged over the  ``background'' region in the sky 
($180^\circ \leq  \alpha \leq 200^\circ$ and  
$20^\circ \leq  \delta \leq 40^\circ$  -- 
a half  of   the scanning area outlined by the dotted line in Fig.~\ref{f:rect})
in which  there are no significant peaks in the power spectrum.

Using the set of recipes presented above, 
we get the following results:    \\
1.\   In all cases considered in  Figs. \ref{f:PXk},\   \ref{f:cuboid} and \ref{f:model}(c)
we got the same significant periods $116 \pm 10~h^{-1}$~Mpc as in Paper~II,
which used  the  data on LRGs of  the  SDSS DR7   and  preliminary  exploited
the data on galaxies of  the  DR12.  \\
2.\  The  special   $X_0$ - directions  considered in this paper 
as well as  in Paper~II     locate  within  a compact region in the sky
$\alpha_0 \simeq 172^\circ \pm 5^\circ$  and  
$\delta_0  \simeq 26^\circ \pm 2^\circ$.  \\
3.\  In the case of a rotating cuboid, when it  gradually captures  the  large  number of 
galaxies  in different  phases  of rotation,  we  
get the highest significance peak  ($ \ga 5 \sigma$) corresponding to the same 
period  $116 \pm 10~h^{-1}$~Mpc  and the same maximum-peak direction $X_0$
as in the cases of  the Radon projections.   \\
4.\   The distribution of the  number density  
of   relatively bright  LOWZ  galaxies
at  the  absolute  $g$-band  magnitude  ${\rm M}_g  \leq - 21.8$, 
lying within the large rectangle in  Fig.~\ref{f:rect},  is  not   homogeneous
relative to the comoving distances  $D \la 1000~h^{-1}$~Mpc\  ($z \la 0.36$), 
but  rather corresponds to a linear regression, 
i.e. strictly speaking it  is not  the  volume-limited sample. 
The  number density  approaches to  homogeneous  one  only for 
the brightest and rarest objects  ${\rm M}_g \leq -22.2$ 
for which  the statistics are  small.   \\
5.\    The peaks in the power spectra of 1D distributions of the projections 
of the  Cartesian coordinates of galaxies  on the special  $X_0$-axis  
turn out to be resistant to the  replenishment  
by  a large number of random  points  (artificial galaxies)
in order to  reach  the  homogeneous total number density. 
The position of the peaks shifts somewhat (within  errors), 
and their high significance is  retained.     \\
6.\   The  estimations produced for the sample of LOWZ galaxies
providing projections onto 
the  $X_0$ direction  in Fig.~\ref{f:PXk}   
show   that  a  signal-to-noise ratio
for   the  major  spatial   harmonic  (at   $k=k_{\rm max}$) 
is   $\ga 3.0$,
then  the  contrast of  1D  density 
projections of  coordinates of galaxies   $N_X(X)$
corresponding to the same harmonic
can be estimated  as   $\ga  0.1$.%
\footnote{    
Following Paper~II, 
we use a modification of  Eqs.~(6),  (7) and (9) 
of  \citet{sc82},
which we  separately  tested  with  simulations.
It  allows  to  estimate  the amplitude of the  fundamental  harmonic 
$A(k_{\rm max})$ included in  the signal-to-noise  ratio
(Eq.~\ref{NNX}), i.e.  
$\langle S/N \rangle (k_{\rm max}) = A(k_{\rm max}) \ga  
\langle \sqrt{4 P(k_{\rm max})/ {\cal N}_b} \rangle$,   
where $P(k_{\rm max})$ is a  height  
of the main peaks in the power spectra, 
${\cal N}_b$  is  a  number of bins  accepted in  each direction $X$,
$\langle ... \rangle$ is  averaging  over all bins along  $X_0$-axis. 
Then  1D  density contrast corresponding to the  fundamental  harmonic  is 
$\delta (k_{\rm max}) \ga  \sqrt{ 4 P(k_{\rm max})/ N_G }$,
where $N_G = {\cal N}_b \cdot S_X$ 
is a  full number of   projections  of coordinates  
onto  the  $X_0$ axis,   $S_X$  is  the average number of 
 the  projections  per a bin.}
Worth to note  that  these characteristics were obtained 
only when applying the method of projections onto the X axis, 
i.e. essentially an integral method that collects information from 
a large amount of  galaxies scattered over the space. 
In reality, we are dealing 
with an extremely  weak and  shallow  ripples  
in the spatial distribution of galaxies,
in contrast to  much more prominent quasi-regular formations 
found in the spatial distribution of galaxy {\it superclusters} at $z \la 0.13$\ 
(e.g.\  \citealt{nat_einast97},\  \citealt{einast97b},\  \citealt{einast14},\  
\citealt{saar02}).

All  conclusions  of this work,  made  so far,  are
obtained on the basis of a statistical analysis of the spectroscopic
measurement of redshifts  $z$, 
at which the accuracy of determination  $z  < 0.5$ is  approximately 
$\delta z \approx 4 \times 10^{-4}$    (e.g. \citealt{bolton12}). 
This is quite sufficient accuracy to identify the  average  period 
$\Delta z \sim 0.04$  over  the  entire  interval $z$. 
However, 
over the past decades  such statistical methods have appeared, 
associated with huge amounts of data, 
which have made it possible to significantly refine 
the photometric measurements of
$z$ (e.g.   \citealt{devicente16} and references therein) for
both galaxies and clusters of galaxies, the latter with greater accuracy
(e.g.  \citealt{wh21, wh22}).  
 
Trying to expand the  scope of  the  search
we carried out a series of very  preliminary calculations 
to study the spatial distributions of galaxy  clusters with a relatively 
accurate definition of   photometric $z$
(results will be published elsewhere).  In particular,
we analyzed the spatial distribution of clusters,
located in the {\it southern} part of the sky, based on the catalog,
presented by  \citet{wh22}. The uncertainty of the cluster 
redshifts $z$  is   $\la 0.013$,
which is quite sufficient to identify the quasi-periodicity  
pertaining to  the  peaks in the power spectra at 
 $0.04 < k_{\rm max} < 0.06$.\  The interval is 
slightly shifted with respect to the LOWZ 
spectroscopic data analysis.

\begin{figure}     
\includegraphics[width=0.45\textwidth]{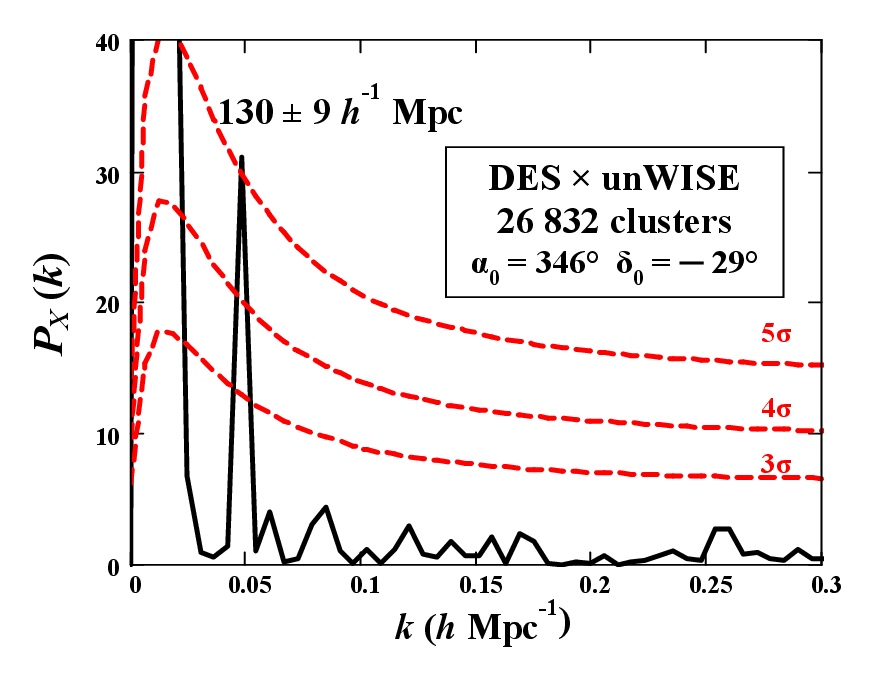}%
\caption{
(Colour online) 
1D power spectrum $P_X(k)$ 
({\it solid  curve}), 
calculated for the distribution of projections of the Cartesian coordinates 
of galaxy  clusters on the $X_0$-axis corresponding to
maximum peak height at    $0.04 <  k_{\rm max} < 0.06~h~{\rm Mpc}^{-1}$
({\it similar to the right panel in} Fig.~\protect{\ref{f:PXk}}).
The spectrum was  calculated  for an array of  clusters from the catalog
DES$\times$unWISE 
with $z$  determined from {\it photometric} data 
with accuracy $\delta z \la 0.013 $.  
Dashed curves are drawn according to the significance levels of the peaks
($3 \sigma, 4\sigma$, and $5\sigma$,  respectively) 
extended to the entire considered range of $k$;
the maximum peak corresponds to 
$k_{\rm max} = 0.048 \pm 0.004~h~{\rm Mpc}^{-1}$.
}  
\label{f:south}
\end{figure}
We have  applied the technique described in 
Sects.~\ref{sec:bd} and \ref{sec:rectang} 
to carry out analysis  of   $26,832$  
clusters from the  catalog,%
\footnote{http://zmtt.bao.ac.cn/galaxy$\_$clusters/catalogs.html} 
\  (file:\   cluster$\_$DESunWISE.dat.gz)\   
with photometric $z$  belonging to  an interval   $0.1 \leq z  \leq 0.47$.

Based on this statistics we  built
the Radon transforms   and  using\  Eq.~(\ref{PXk}) 
calculated  the power  spectra  of 
1D   distributions  of  projections  of  cluster  coordinates 
on the $X$-axes  with   different orientations,   
but  at  fixed  interval
$132 \leq  X  \leq 1172~h^{-1}$~Mpc\  
($0.1 \leq z \leq 0.47$). 
Directions of $X$-axis  were  changed 
within  an angular  area 
$340^\circ \leq   \alpha \leq  360^\circ$,\quad   
$-40^\circ \leq  \delta  \leq  -20^\circ$  
with a step by one degree. 
The same region of angular scanning 
with the $X$-axis was used in order 
to calculate the  significance levels for all considered $k$
($0 < k < 0.3~h~{\rm Mpc}^{-1})$.

Fig.~\ref{f:south} shows the preliminary results of such calculations. 
One can see a significant dominant peak at  $k = 0.048 \pm 0.004~h~{\rm Mpc}^{-1}$
(period $130 \pm 9~h^{-1}$~Mpc)
in the 1D  power spectrum corresponding to the $X_0$ direction
with angular coordinates 
$\alpha_0 \simeq 346^\circ$ 
and $\delta_0 \simeq -29^\circ$. 
Significance  of the peak  exceeded  $5\sigma$.

Thus the  indicated  direction 
is approximately a continuation 
of the selected direction 
in the {\it northern} hemisphere (see above),
which is reminiscent of  the  results 
by  \citet{beks90}, \citet{szetal93},
\citet{koo93} (see discussion in Paper~II).  
Although the selected bundle   of  close directions 
(with high level of  significance of the peaks)
is somewhat rotated  (by  $\sim  20^\circ - 30^\circ$) 
in  right ascension $\alpha$  relative to  the axis   connecting 
the north  and south  galactic poles
$(\alpha_{\rm ngp} =192.85^\circ,  \quad
\alpha_{\rm sgp} = 12.85^\circ (372.85^\circ)).$
In the same time  the declinations  $\delta$ 
are close  the declinations  
of {\it northern} ($\delta_{\rm ngp}=27.13^\circ$)
and {\it southern} ($\delta_{\rm sgp} = -27.13^\circ$) 
poles, respectively.
Consequently, we got preliminary sketches of  weak
large-scale anisotropy in the distribution of 
galaxies(north) and   clusters(south), 
which may have a dipole character  with slightly different
features  in the  north and south  sky.
This fact in itself requires careful verification and further 
confirmation.

Let us  note, that  some small  discrepancy between 
north and south data 
was found recently by  calculations 
of  the Minkowski  functionals (\citealt{appbuch22})
on the basis of the  LOWZ statistics. 
These results do not contradict  to
the assumption of  the  anisotropy 
of  the spatial distribution of matter  (emphasized  by the authors), 
as well as  the assumption  about 
large scale inhomogeneity of the distribution, 
since the data of the LOWZ 
cover a limited  domain  of space.
While
the calculations based on the CMASS data 
(corresponding to  larger  redshifts)
no longer show north/south discrepancy 
(see \citealt{appbuch22} for details).

Thus the present study 
confirms  the hypothesis formulated in Paper~II  that 
at the redshifts considered here,
there may be a huge elongated quasi-periodic
structure that  represents  an alternation of flat
condensation and rarefaction of matter
with a characteristic scale
$\Delta X = 116 \pm 10~ h^{-1}$~Mpc.
Along  the directions close to that found in this work
the structure is likely to have a  total   
scale  $\ga  800~h^{-1}$~Mpc. 
In other words we find 
traces  of two scales 
of an anisotropic quasi-regular structure
the  quasi-period pointed  out  above  and entire  scale, on which  
this period  appears.
At the same time, the very existence of such a structure
remains a hypothesis requiring further confirmation.

\section*{Acknowledgments}
We are extremely grateful to an anonymous reviewer for 
his numerous and very helpful comments.

\section*{Data availability}
%
%
[dataset]* Reid B. et al.,\ 2016,\  Large-scale structure catalogues,\\  
galaxy$\_$DR12v5$\_$LOWZ$\_$North.fits.gz,\\ 
galaxy$\_$DR12v5$\_$ CMASSLOWZE3 $\_$North.fits.gz,\\ 
galaxy$\_$DR12v5$\_$CMASS $\_$North.fits.gz,\\ 
https://data.sdss3.org/sas/dr12/boss/lss/   \\ 
\par
\noindent
[dataset]* Beck R. et al.,\ 2016,\  SDSS  SkyServey,\    Sample   SQL  Queries   ,\
http://skyserver.sdss.org/dr12/en/help/docs/realquery.aspx  \\
\par
\noindent
[dataset]* Kazin E.A. et al.,\  2010,\    Sloan-digital sky survey LRG sample,\   Full DR7 LRG sample (data ascii),\
https://cosmo.nyu.edu/$\sim$eak306/SDSS-LRG.html    \\
\par
\noindent
[dataset]*  Wen Z.L., Han J.L.,\  2022,\  Catalogs of clusters of galaxies,\   cluster$\_$DESunWISE.dat.gz,\                      
http://zmtt.bao.ac.cn/galaxy$\_$clusters/catalogs.html        \\

\bibliographystyle{mnras}
\bibliography{BibList-LSS}

\begin{thebibliography}{}
\makeatletter
\relax
\def\mn@urlcharsother{\let\do\@makeother \do\$\do\&\do\#\do\^\do\_\do\%\do\~}
\def\mn@doi{\begingroup\mn@urlcharsother \@ifnextchar [ {\mn@doi@}
  {\mn@doi@[]}}
\def\mn@doi@[#1]#2{\def\@tempa{#1}\ifx\@tempa\@empty \href
  {http://dx.doi.org/#2} {doi:#2}\else \href {http://dx.doi.org/#2} {#1}\fi
  \endgroup}
\def\mn@eprint#1#2{\mn@eprint@#1:#2::\@nil}
\def\mn@eprint@arXiv#1{\href {http://arxiv.org/abs/#1} {{\tt arXiv:#1}}}
\def\mn@eprint@dblp#1{\href {http://dblp.uni-trier.de/rec/bibtex/#1.xml}
  {dblp:#1}}
\def\mn@eprint@#1:#2:#3:#4\@nil{\def\@tempa {#1}\def\@tempb {#2}\def\@tempc
  {#3}\ifx \@tempc \@empty \let \@tempc \@tempb \let \@tempb \@tempa \fi \ifx
  \@tempb \@empty \def\@tempb {arXiv}\fi \@ifundefined
  {mn@eprint@\@tempb}{\@tempb:\@tempc}{\expandafter \expandafter \csname
  mn@eprint@\@tempb\endcsname \expandafter{\@tempc}}}

\bibitem[\protect\citeauthoryear{{Alam} et~al.,}{{Alam} et~al.}{2015}]{alam15}
{Alam} S.,  et~al., 2015, \apjs, 219, 12

\bibitem[\protect\citeauthoryear{{Appleby}, {Park}, {Pranav}, {Hong}, {Hwang},
  {Kim}  \& {Buchert}}{{Appleby} et~al.}{2022}]{appbuch22}
{Appleby} S.,  {Park} C.,  {Pranav} P.,  {Hong} S.~E.,  {Hwang} H.~S.,  {Kim}
  J.,   {Buchert} T.,  2022, \mn@doi [\apj] {10.3847/1538-4357/ac562a}, 928,
  108

\bibitem[\protect\citeauthoryear{{Bardeen}, {Bond}, {Kaiser}  \&
  {Szalay}}{{Bardeen} et~al.}{1986}]{bbks86}
{Bardeen} J.~M.,  {Bond} J.~R.,  {Kaiser} N.,   {Szalay} A.~S.,  1986, \apj,
  304, 15

\bibitem[\protect\citeauthoryear{{Beck}, {Dobos}, {Budav{\'a}ri}, {Szalay}  \&
  {Csabai}}{{Beck} et~al.}{2016}]{beck16}
{Beck} R.,  {Dobos} L.,  {Budav{\'a}ri} T.,  {Szalay} A.~S.,   {Csabai} I.,
  2016, \mn@doi [\mnras] {10.1093/mnras/stw1009}, 460, 1371

\bibitem[\protect\citeauthoryear{{Bolton} et~al.,}{{Bolton}
  et~al.}{2012}]{bolton12}
{Bolton} A.~S.,  et~al., 2012, \aj, 144, 144

\bibitem[\protect\citeauthoryear{{Broadhurst}, {Ellis}, {Koo}  \&
  {Szalay}}{{Broadhurst} et~al.}{1990}]{beks90}
{Broadhurst} T.~J.,  {Ellis} R.~S.,  {Koo} D.~C.,   {Szalay} A.~S.,  1990,
  \mn@doi [\nat] {10.1038/343726a0}, \href
  {https://ui.adsabs.harvard.edu/abs/1990Natur.343..726B} {343, 726}

\bibitem[\protect\citeauthoryear{{Dawson} et~al.,}{{Dawson}
  et~al.}{2013}]{dawson13}
{Dawson} K.~S.,  et~al., 2013, \mn@doi [\aj] {10.1088/0004-6256/145/1/10}, 145,
  10

\bibitem[\protect\citeauthoryear{{De Vicente}, {S{\'a}nchez}  \&
  {Sevilla-Noarbe}}{{De Vicente} et~al.}{2016}]{devicente16}
{De Vicente} J.,  {S{\'a}nchez} E.,   {Sevilla-Noarbe} I.,  2016, \mn@doi
  [\mnras] {10.1093/mnras/stw857}, 459, 3078

\bibitem[\protect\citeauthoryear{{Deans}}{{Deans}}{2007}]{deans07}
{Deans} S.~R.,  2007, {The Radon transform and some of its applications}.
New York, Dover Publications., Inc.

\bibitem[\protect\citeauthoryear{{Einasto}}{{Einasto}}{2014}]{einast14}
{Einasto} J.,  2014, {Dark Matter and Cosmic Web Story}.
World Scientific

\bibitem[\protect\citeauthoryear{{Einasto}, {Einasto}, {Tago}, {Dalton}  \&
  {Andernach}}{{Einasto} et~al.}{1994}]{einastM94}
{Einasto} M.,  {Einasto} J.,  {Tago} E.,  {Dalton} G.~B.,   {Andernach} H.,
  1994, \mnras, 269, 301

\bibitem[\protect\citeauthoryear{{Einasto}, {Tago}, {Jaaniste}, {Einasto}  \&
  {Andernach}}{{Einasto} et~al.}{1997a}]{einast97a}
{Einasto} M.,  {Tago} E.,  {Jaaniste} J.,  {Einasto} J.,   {Andernach} H.,
  1997a, \aaps, 123, 119

\bibitem[\protect\citeauthoryear{{Einasto} et~al.,}{{Einasto}
  et~al.}{1997b}]{einast97b}
{Einasto} J.,  et~al., 1997b, \mnras, 289, 801

\bibitem[\protect\citeauthoryear{{Einasto}, {Einasto}, {Frisch}, {Gottlober},
  {Muller}, {Saar}, {Starobinsky}  \& {Tucker}}{{Einasto}
  et~al.}{1997c}]{einast97c}
{Einasto} J.,  {Einasto} M.,  {Frisch} P.,  {Gottlober} S.,  {Muller} V.,
  {Saar} V.,  {Starobinsky} A.~A.,   {Tucker} D.,  1997c, \mnras, 289, 813

\bibitem[\protect\citeauthoryear{{Einasto} et~al.,}{{Einasto}
  et~al.}{1997d}]{nat_einast97}
{Einasto} J.,  et~al., 1997d, \nat, 385, 139

\bibitem[\protect\citeauthoryear{{Einasto} et~al.,}{{Einasto}
  et~al.}{2011a}]{einast11}
{Einasto} J.,  et~al., 2011a, \aap, 531, A75

\bibitem[\protect\citeauthoryear{{Einasto} et~al.,}{{Einasto}
  et~al.}{2011b}]{einast11a}
{Einasto} J.,  et~al., 2011b, \mn@doi [\aap] {10.1051/0004-6361/201117248},
  534, A128

\bibitem[\protect\citeauthoryear{{Einasto} et~al.,}{{Einasto}
  et~al.}{2016}]{einastM16_shell}
{Einasto} M.,  et~al., 2016, Astron. Astrophys., 587, A116

\bibitem[\protect\citeauthoryear{{Einasto}, {Suhhonenko}, {Liivam{\"a}gi}  \&
  {Einasto}}{{Einasto} et~al.}{2019}]{einast19}
{Einasto} J.,  {Suhhonenko} I.,  {Liivam{\"a}gi} L.~J.,   {Einasto} M.,  2019,
  \mn@doi [\aap] {10.1051/0004-6361/201834450}, 623, A97

\bibitem[\protect\citeauthoryear{{Feldman}, {Kaiser}  \& {Peacock}}{{Feldman}
  et~al.}{1994}]{fkp94}
{Feldman} H.~A.,  {Kaiser} N.,   {Peacock} J.~A.,  1994, \apj, 426, 23

\bibitem[\protect\citeauthoryear{{Hogg}}{{Hogg}}{1999}]{h99}
{Hogg} D.~W.,  1999, astro-ph/9905116

\bibitem[\protect\citeauthoryear{{Kaiser} \& {Peacock}}{{Kaiser} \&
  {Peacock}}{1991}]{kp91}
{Kaiser} N.,  {Peacock} J.~A.,  1991, \apj, 379, 482

\bibitem[\protect\citeauthoryear{{Kayser}, {Helbig}  \& {Schramm}}{{Kayser}
  et~al.}{1997}]{khs97}
{Kayser} R.,  {Helbig} P.,   {Schramm} T.,  1997, Astron. Astrophys., 318

\bibitem[\protect\citeauthoryear{{Kazin} et~al.,}{{Kazin}
  et~al.}{2010}]{kazetal10}
{Kazin} E.~A.,  et~al., 2010, \apj, 710, 1444

\bibitem[\protect\citeauthoryear{{Kerscher}}{{Kerscher}}{1998}]{kerscher98}
{Kerscher} M.,  1998, \aap, 336, 29

\bibitem[\protect\citeauthoryear{{Kerscher}}{{Kerscher}}{2000}]{kerscher00}
{Kerscher} M.,  2000, in {Mecker} K.~R.,  {Stoyan} D.,  eds,  Lecture Notes in
  Physics Vol. 554, Statistical Physics and Spatial Statistics. Berlin Springer
  Verlag, pp 36--71

\bibitem[\protect\citeauthoryear{{Kerscher} et~al.,}{{Kerscher}
  et~al.}{1997}]{kerscher97}
{Kerscher} M.,  et~al., 1997, \mn@doi [\mnras] {10.1093/mnras/284.1.73}, 284,
  73

\bibitem[\protect\citeauthoryear{{Koo}, {Ellman}, {Kron}, {Munn}, {Szalay},
  {Broadhurts}  \& {Ellis}}{{Koo} et~al.}{1993}]{koo93}
{Koo} D.~C.,  {Ellman} N.,  {Kron} R.~G.,  {Munn} J.~A.,  {Szalay} A.~S.,
  {Broadhurts} T.~J.,   {Ellis} R.~S.,  1993, in {Chincarini} G.~L.,  {Iovino}
  A.,  {Maccacaro} T.,   {Maccagni} D.,  eds,  Astronomical Society of the
  Pacific Conference Series Vol. 51, Observational Cosmology. p.~112

\bibitem[\protect\citeauthoryear{{Landy}, {Shectman}, {Lin}, {Kirshner},
  {Oemler}  \& {Tucker}}{{Landy} et~al.}{1996}]{landy96}
{Landy} S.~D.,  {Shectman} S.~A.,  {Lin} H.,  {Kirshner} R.~P.,  {Oemler}
  A.~A.,   {Tucker} D.,  1996, \apjl, 456, L1

\bibitem[\protect\citeauthoryear{{Ledermann} \& {Lloyd}}{{Ledermann} \&
  {Lloyd}}{1984}]{ll84}
{Ledermann} W.,  {Lloyd} E.,  1984, {Handbook of Applicable Mathematics. Vol.
  VI: Statistics}.
John Wiley \& Sons, New York

\bibitem[\protect\citeauthoryear{{Mecke}}{{Mecke}}{2000}]{mecke00}
{Mecke} K.~R.,  2000, in {Mecker} K.~R.,  {Stoyan} D.,  eds,  Lecture Notes in
  Physics Vol. 554, Statistical Physics and Spatial Statistics. Berlin Springer
  Verlag, pp 111--184

\bibitem[\protect\citeauthoryear{{Reid} et~al.,}{{Reid} et~al.}{2016}]{reid16}
{Reid} B.,  et~al., 2016, \mnras, 455, 1553

\bibitem[\protect\citeauthoryear{{Ryabinkov} \& {Kaminker}}{{Ryabinkov} \&
  {Kaminker}}{2014}]{rk14}
{Ryabinkov} A.~I.,  {Kaminker} A.~D.,  2014, \mnras, 440, 2388

\bibitem[\protect\citeauthoryear{{Ryabinkov} \& {Kaminker}}{{Ryabinkov} \&
  {Kaminker}}{2019}]{rk19}
{Ryabinkov} A.~I.,  {Kaminker} A.~D.,  2019, \apss, 364, 129\, \, (Paper~I)

\bibitem[\protect\citeauthoryear{{Ryabinkov} \& {Kaminker}}{{Ryabinkov} \&
  {Kaminker}}{2021}]{rk21}
{Ryabinkov} A.~I.,  {Kaminker} A.~D.,  2021, Universe, 7, 289\, \, (Paper~II)

\bibitem[\protect\citeauthoryear{{Ryabinkov}, {Kaurov}  \&
  {Kaminker}}{{Ryabinkov} et~al.}{2013}]{rkk13}
{Ryabinkov} A.~I.,  {Kaurov} A.~A.,   {Kaminker} A.~D.,  2013, \apss, 344, 219

\bibitem[\protect\citeauthoryear{{Saar}, {Einasto}, {Toomet}, {Starobinsky},
  {Andernach}, {Einasto}, {Kasak}  \& {Tago}}{{Saar} et~al.}{2002}]{saar02}
{Saar} E.,  {Einasto} J.,  {Toomet} O.,  {Starobinsky} A.~A.,  {Andernach} H.,
  {Einasto} M.,  {Kasak} E.,   {Tago} E.,  2002, \aap, 393, 1

\bibitem[\protect\citeauthoryear{{Sahni}, {Sathyaprakash}  \&
  {Shandarin}}{{Sahni} et~al.}{1998}]{sahni98}
{Sahni} V.,  {Sathyaprakash} B.~S.,   {Shandarin} S.~F.,  1998, \mn@doi [\apjl]
  {10.1086/311214}, 495, L5

\bibitem[\protect\citeauthoryear{{Scargle}}{{Scargle}}{1982}]{sc82}
{Scargle} J.~D.,  1982, \apj, 263, 835

\bibitem[\protect\citeauthoryear{{Starck}, {Mart{\'\i}nez}, {Donoho}, {Levi},
  {Querre}  \& {Saar}}{{Starck} et~al.}{2005}]{starcketal05}
{Starck} J.~L.,  {Mart{\'\i}nez} V.~J.,  {Donoho} D.~L.,  {Levi} O.,  {Querre}
  P.,   {Saar} E.,  2005, EURASIP Journal on Applied Signal Processing, p.
  483071

\bibitem[\protect\citeauthoryear{{Szalay}, {Ellis}, {Koo}  \&
  {Broadhurst}}{{Szalay} et~al.}{1991}]{szetal91}
{Szalay} A.~S.,  {Ellis} R.~S.,  {Koo} D.~C.,   {Broadhurst} T.~J.,  1991, in
  {Holt} S.~S.,  {Bennett} C.~L.,   {Trimble} V.,  eds,  American Institute of
  Physics Conference Series Vol. 222, After the first three minutes. p.~261

\bibitem[\protect\citeauthoryear{{Szalay}, {Broadhurst}, {Ellman}, {Koo}  \&
  {Ellis}}{{Szalay} et~al.}{1993}]{szetal93}
{Szalay} A.~S.,  {Broadhurst} T.~J.,  {Ellman} N.,  {Koo} D.~C.,   {Ellis}
  R.~S.,  1993, Proc. Natl. Acad. Sci. USA, 90, 4853

\bibitem[\protect\citeauthoryear{{Wen} \& {Han}}{{Wen} \& {Han}}{2021}]{wh21}
{Wen} Z.~L.,  {Han} J.~L.,  2021, \mn@doi [\mnras] {10.1093/mnras/staa3308},
  500, 1003

\bibitem[\protect\citeauthoryear{{Wen} \& {Han}}{{Wen} \& {Han}}{2022}]{wh22}
{Wen} Z.~L.,  {Han} J.~L.,  2022, \mn@doi [\mnras] {10.1093/mnras/stac1149},
  513, 3946

\bibitem[\protect\citeauthoryear{{Wen}, {Han}  \& {Liu}}{{Wen}
  et~al.}{2012}]{whl12}
{Wen} Z.~L.,  {Han} J.~L.,   {Liu} F.~S.,  2012, \apjs, 199, 34

\bibitem[\protect\citeauthoryear{{Yoshida} et~al.,}{{Yoshida}
  et~al.}{2001}]{yoshida01}
{Yoshida} N.,  et~al., 2001, \mnras, 325, 803

\bibitem[\protect\citeauthoryear{{van de Weygaert}}{{van de
  Weygaert}}{2016}]{weygaert16}
{van de Weygaert} R.,  2016, in {van de Weygaert} R.,  {Shandarin} S.,  {Saar}
  E.,   {Einasto} J.,  eds,  IAU Symposium Vol. 308, The Zeldovich Universe:
  Genesis and Growth of the Cosmic Web. pp 493--523

\bibitem[\protect\citeauthoryear{{van de Weygaert} \& {Schaap}}{{van de
  Weygaert} \& {Schaap}}{2009}]{weygaert09}
{van de Weygaert} R.,  {Schaap} W.,  2009, in {Mart{\'\i}nez} V.~J.,  {Saar}
  E.,  {Mart{\'\i}nez-Gonz{\'a}lez} E.,   {Pons-Border{\'\i}a} M.~J.,  eds,
  Lecture Notes in Physics Vol. 665, Data Analysis in Cosmology. Berlin
  Springer Verlag, pp 291--413

\makeatother
\end{thebibliography}

\end{document}